\documentclass[11pt]{article}
\usepackage{jheppub}
\usepackage{pstool}
\pdfoutput=1
\usepackage{color}
\usepackage{float}
\usepackage{array}
\usepackage{amssymb}
\usepackage{amsmath}
\usepackage{amsthm}
\usepackage{graphicx}
\usepackage{caption}
\usepackage[labelsep=quad]{subcaption}
\usepackage{epstopdf}
\usepackage{placeins}
	
\usepackage{epsfig}
\usepackage{simpler-wick}

\def\pb[#1,#2]{\{#1, #2\}}
\def\deb[#1,#2]{[#1,#2]_{\text{D.B.}}}

\def\Or[#1]{{\text{O}}\left({#1}\right)}
\def\dotl[#1,#2]{\left\langle #1,\, #2 \right\rangle}
\def\dotlb[#1,#2]{\left\langle #1,\, #2 \right\rangle}
\def\dotlm[#1,#2]{\left[ #1,\, #2 \right]}
\def\dotp[#1,#2]{(\vect{#1} \cdot\vect{#2})}
\def\aff[#1,#2]{\hat{#1}(#2)}

\def\n4sym{{\cal N}=4 SYM}
\def\>{\rangle}
\def\<{\langle}
\def\weight[#1,#2,#3]{\{(#1),#2,#3\}}
\def\ads[#1]{$\text{AdS}_{#1}$}

\newcommand{\be}{\begin{equation}}
\newcommand{\ee}{\end{equation}}
\newcommand{\ba}{\begin{align}}
\newcommand{\ea}{\end{align}}
\newcommand{\bs}{\begin{split}}
	\def\sess\end{split}
\newcommand{\vect}[1]{{\boldsymbol{#1}}}

\def \bea {\begin{eqnarray}}
\def \eea {\end{eqnarray}}
\def \bea* {\begin{eqnarray*}}
	\def \eea* {\end{eqnarray*}}

\def \be {\begin{equation}}
\def \ee {\end{equation}}
\def \bes {\begin{equation*}}
\def \ees {\end{equation*}}

\title{Coarse-graining in time, emergent temperature in 2d CFT at large central charge}
\author[a]{Gideon Vos}
\emailAdd{gideonvos@kias.re.kr}
\affiliation[a]{School of Physics, Korea Institute for Advanced Study, 85 Hoegi-ro, Dongdaemun-gu, Seoul 02455, Republic of Korea}

\date{}

\abstract{Using the monodromy method, a compact expression is obtained for the identity block contribution to the expectation value of two low-energy probe operators on a broad class of time-dependent heavy pure states in large-$c$ 2d CFTs. It will be shown that the size of the compact spatial dimension sets a coarse-graining time-scale above which the thermal two-point function naturally emerges. A phenomena will be highlighted where the information of the pure state appears to get projected down to a single conformal representation. We will see that this projection is corroborated by known conformal bootstrap results.}
\begin{document}
	\maketitle
	


\section{Introduction}
The domain of two-dimensional conformal field theories is situated on an interesting cusp. Their rigid symmetry structure allows us to write down exact analytic expressions for correlation functions at strong coupling, the density of states in the deep UV \cite{CARDY1986186} or the entanglement entropy of a subregion \cite{Calabrese:2009qy}. Quantities that are at current capabilities near-impossible to compute in generic quantum field theories. Despite this two-dimensional CFTs are sufficiently unconstrained to display a wide range of phenomena. In some cases they are integrable \cite{Bazhanov:1994ft}, in some cases they display complex out-of-equilibrium phenomena such such as thermalization \cite{Balasubramanian:2010ce,Balasubramanian:2011ur,deBoer:2016bov} or quantum chaos \cite{Roberts:2014ifa}. These combined properties make this class of models excellent laboratories for emergent thermal physics.

The level of simplification is further enhanced by assuming the central charge $c$ is very large. Heuristically the central charge counts the number of degrees of freedom of the system and acts analogously to a large-$N$ limit. Motivated by holography, where large central charge corresponds to semi-classical bulk gravity, this has resulted in a large body of literature \cite{Fitzpatrick:2014vua, Fitzpatrick:2015zha, Martinec:1998wm,Perlmutter:2015iya,Anous:2016kss, Balasubramanian:2017fan,Alkalaev:2017bzx, Bissi:2024wur}, including a large body of literature on entanglement entropy at large central charge \cite{Hartman:2013mia,Asplund:2014coa, Chen:2016dfb,Chen:2016kyz}.



Though without consideration of holography the large large-$c$ regime, possesses interesting aspects of collective many-body phenomena. This paper will explore some of those aspects purely from a CFT perspective by considering a broad class of time-dependent high energy pure states. Some of the underlying physics of this state will be extracted by considering the expectation value of two low-energy probe observables on this high-energy state. In particular the behavior of this expectation value as the two light operators are separated in time will be studied.


\subsection{Outline and summary}
\begin{figure}
	\centering
		\includegraphics[scale=1]{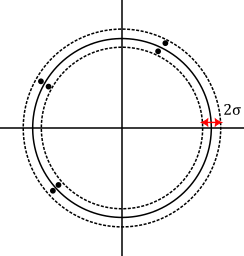}
	\caption{Restricting the heavy operator to lie within a thin annulus of width $2\sigma$ centered around the unit circle gives us a controllable parameter with which the identity block dominance of the Euclidean correlator can be controlled.}
	\label{annulusintroduction}
\end{figure}

The framework under consideration consists of 2d CFT with large central charge $c\gg1$. Under this assumption operators can be sorted into two non-exhaustive hierarchies
\begin{align}
    & \text{Heavy operators:} \;\;\; [L_0,O_H] = HO_H, \;\;\;\;\;\; \frac{H}{c}\propto \mathcal{O}(c^0), \\
    & \text{Light operators:} \;\;\;\; [L_0,O_h] = hO_h, \;\;\;\;\;\;\;\;\; \frac{h}{c} \propto \mathcal{O}(1/c), 
\end{align}
Where $L_0$ is zeroth Virasoro generator. The object of interest this work will be a class of high-energy CFT states $|V\rangle$. This class of states will be created by acting with a \textit{generic} (though much smaller than $c$) number of heavy operators on the vacuum state
\begin{equation}
  |V\rangle =  O_H(x_1)...O_H(x_n)|0\rangle.
\label{heavystate}
\end{equation}
These states by construction are time-dependent. In order to extract information from the heavy state the normalized expectation value of two light operators will be constructed
\begin{equation}
    \frac{\langle V|O_L(z_1)O_L(z_2)|V\rangle}{\langle V|V\rangle}.
\label{expectionvalueintroduction}
\end{equation}
The points $x_i$ of the heavy operator location in \eqref{heavystate} will not be completely generic. The following analysis borrows heavily from the structure of \cite{Banerjee:2018tut}, but refines the discussion on a number of points. For practical reasons they will be assumed to lie in a thin annulus of width $2\sigma$ surrounding the unit circle, see figure \ref{annulusintroduction}. This ensures that the OPE channel where a heavy operator $O_H(x_i)$ in $|V\rangle$ OPE contracted with its adjoint image $O_H^{\dagger}(1/\bar{x}_i)$ is dominated by the identity block exchange. 

The identity block contribution will be computed using the monodromy method \cite{BELAVIN1984333, Hartman:2013mia, Fitzpatrick:2014vua}. The main novel analytical contribution of this work is a compact formula for the light operator dependence of the identity block of \eqref{expectionvalueintroduction}. Dubbing this light sector of the conformal block $Q(z_1,z_2)$, it is given by
\begin{equation}
    \boxed{Q(z_1,z_2) = \frac{1}{\left(\rho_+(z_1)\rho_-(z_2) - \rho_+(z_2)\rho_-(z_1)\right)^{2h}}},
\label{mainresult}
\end{equation}
where $\rho_{\pm}(z)$ is a Wronskian normalized\footnote{i.e. $\rho_+(z)\rho_-'(z)-\rho_-(z)\rho_+'(z) =1$.} linearly independent basis of solutions of the 2nd order ODE
\begin{equation}
    \rho''_{\pm}(z) + \frac{6}{c}T_H(z)\rho_{\pm}(z)=0,
\label{Fuchsianintroduction}
\end{equation}
where the potential $T_H(z)$ is given by the normalized expectation value of the stress tensor of the sector of heavy operators
\begin{equation}
    T_H(z)\equiv \frac{\langle V|T(z)|V\rangle}{\langle V|V\rangle}.
\end{equation}
Owing to the generality of the states \ref{heavystate}, the basis of solutions is very difficult to construct, but it has enough structure that some interesting physics can be extracted. The expression \eqref{mainresult} has a number of appealing features, it is relatively compact, it is manifestly independent of choice of basis of solutions of \eqref{Fuchsianintroduction}, it reproduces some known results in the literature \cite{Fitzpatrick:2015zha, Banerjee:2018tut}, and in the limit $z_2\rightarrow z_1$ it reduces to the conformal two-point function. In section \ref{Lorentziantime} analytic continuation to Lorentzian time will be considerd, inspired by \cite{Kundu:2025jsm} it will be seen that the branch structure of the solutions $\rho_{\pm}(z)$ exactly mimics the lightcone structure on the Minkowski diagram (see figure \ref{lightcones}).

It will be shown in section \ref{Floquetsection} that the differential equations have enough structure that after analytic continuation in some cases the general structure of the light operator expectation value can be obtained. In particular it will be found that in the case that the solutions to the ODE \eqref{Fuchsianintroduction} are exponential that the light sector of the expectation value becomes thermal if the time-separation of the light operators is taken to be long compared to the size of the compact space dimension. As a result the cylinder size sets a natural coarse-graining scale after which thermal physics appears. 

Finally a small puzzle will be discussed in section \ref{thermodynamicsandorbits}, the state $|V\rangle$ a priori is a linear combination of CFT states belonging to many different conformal families. In the process of assuming large $c$, identity block dominance and analytically continuing to Lorentzian time it was found that the data of $|V\rangle$ that enters into the correlation function is a derived function of time $\mathcal{V}(t)$ that was constructed out $T_H(z)$, see equation \eqref{Hillpotential}. This function $\mathcal{V}(t)$ has all the structure of (part of) a Virasoro coadjoint vector. A Virasoro coadjoint vector is an element of a Virasoro coadjoint orbit which is the classical limit of Virasoro representation \cite{WittenCoadjointOrbits, Alekseev:1988ce,Alekseev:1990mp,Balog:1997zz}. Therefore as a consequence of the large-$c$ regime the data of $|V\rangle$ gets projected down to a single Virasoro representation.

This raises a natural question which is; how is a representation selected out of $|V\rangle$? This question is tackled in section \ref{conformalbootstrap} for the specific case of two identical heavy operators
\begin{equation}
    |V\rangle \rightarrow |V_2\rangle = O_H(x_1)O_H(x_2)|0\rangle.
\end{equation}
In this section it is found that observed projection is corroborated by the conformal bootstrap. This was found by utilizing known asymptotic expressions for the averaged OPE coefficients in the high-scaling dimension regime \cite{Das:2017cnv, Cardy:2017qhl, Collier:2019weq}. By analyzing the location of the dominant saddle-point contribution to the conformal cross-channel a conjecture was derived for a formula relating the external scaling dimensions of the heavy operators $H$ to the temperature experienced by the light operators. Finally, some open problems will be discussed.

\section{Probes from the monodromy method}
\label{monodromymethod}
By utilizing the monodromy method a general derivation for the light sector contribution to the semi-classical conformal identity block \eqref{mainresult} will be provided. In order to do this some terminology will need to be introduced. It is generally accepted \cite{Besken:2019jyw,Hartman:2013mia} that when the central charge $c$ is very large $c\gg1$ the conformal blocks take the leading exponential form
\begin{equation}
    \mathcal{F}_{\text{Heavy}}(H,h_p;x) = e^{-\frac{c}{6}f_0(H_i,h_p,x) +\mathcal{O}(c^0)},
\label{heavyblock}
\end{equation}
here $H$ are the collective external holomorphic scaling dimensions, $h_p$ the internal exchange primaries and $x$ the collective conformal cross-ratios. The leading part $f_0(H,h_p,x)$ is the semiclassical block. Here it will be assumed that all the external scaling dimensions are heavy in the sense that the ratios 
\begin{equation}
\frac{H_i}{c} = \mathcal{O}(c^0).
\end{equation}
As a trivial example, note that the heavy two-point function consists of a single conformal block exchange, the identity block, and is given by
\begin{equation}
\langle O(z_1)O(z_2)\rangle = \frac{1}{(z_1-z_2)^{2H}} = e^{-\frac{c}{6}\left(\frac{12H}{c}\log(z_1-z_2)\right)}.
\end{equation}
In this case the semi-classical block can be recognized as
\begin{equation}
    f_0^{2\text{pt.}}(H,0,z_1-z_2) = \frac{12H}{c}\log(z_1-z_2),
\end{equation}
which indeed only depends on the $\mathcal{O}(c^0)$ ratio $H/c$. In this particular example the leading term is exact, but in general there are subleading corrections.

The next terminology to consider is heavy-light perturbation theory. In contrast to the heavy operators above, light operators have scaling dimensions that are subleading at large $c$
\begin{equation}
    \frac{h}{c} =\mathcal{O}(1/c).
\end{equation}
The conformal blocks of the mixed heavy-light correlator $G(x_i,z_i)$ will be considered, where $N$ heavy operators are inserted at the locations $x_i$ and insert $n$ light operators at $z_j$, i.e.
\begin{equation}
    \mathcal{G}(x_j,z_i) = \langle \mathcal{O}_H(x_1)...\mathcal{O}_H(x_N)\mathcal{O}_L(z_1)...\mathcal{O}_L(z_n)\rangle.
\end{equation}
The heavy operators are not considered all to be equal, but any labels distinguishing them will be suppressed. In \textit{heavy-light perturbation theory} the presence of the light operators is considered to have a perturbative effect on the conformal blocks of the heavy operators. In practical terms, going back to \eqref{heavyblock}, the light operator dependence of the conformal block does not manifest itself until the first subleading term
\begin{equation}
    \mathcal{F}_{\text{Heavy-Light}}(H,h,h_p;z_i,x_i) = e^{-\frac{c}{6}f_0(H;z_i) + f_1(H,h;x_i,z_i) + \mathcal{O}(1/c)}.
\label{conformalblock}
\end{equation}
If correction beyond the first subleading correction is ignored then it can be concluded that the conformal block factorizes into an unperturbed heavy factor and a light factor
\begin{equation}
   \mathcal{F}_{\text{Heavy-Light}}(H,h,h_p;z_i,x_i) \sim e^{-\frac{c}{6}f_0(H;z_i)} e^{ f_1(H,h;x_i,z_i)}.
\end{equation}
In the upcoming sections it will be shown  that the complex analytic structure of the conformal block is rigid enough that we can fix the light sector dependence of the light factor completely in terms of data of the heavy sector.

\subsection{Monodromy method}\label{Monodromy method}
The way this will be accomplished is through means of the monodromy method. Before this method will be described there is another object that need to be introduced. The stress tensor expectation value $T(z)$
\begin{equation}
    \langle\hat{T}(z)\mathcal{O}_H(x_1)...\mathcal{O}_H(x_N)\mathcal{O}_L(z_1)...\mathcal{O}_L(z_n)\rangle = T(z) \mathcal{G}(x_j,z_i),
\end{equation}
where the operator $\hat{T}(z)$ is the holomorphic component of the stress tensor. The Virasoro Ward identity fixes the general form of the function $T(z)$ to the form
\begin{equation}
    T(z) = \sum_{j=1}^N \frac{H_j}{(z-x_j)^2} + \frac{\partial_{x_j}\log(\mathcal{G})}{z-x_j} + \sum_{i=1}^n \frac{h_i}{(z-x_i)^2} + \frac{\partial_{z_i}\log(\mathcal{G})}{z-z_i} 
\label{generalstresstensor}
\end{equation}
Given a choice of OPE channel $\mathcal{G}(x_j,z_i)$ can be written as a sum over conformal blocks
\begin{equation}
    \mathcal{G}(x_j,z_i) = \sum_{\vec{n}} C_{12{n_1}}C_{34{n_2}}...\mathcal{F}_{\text{Heavy-Light}}(H,h,h_p;z_i,x_i),
    \label{blockexpansion}
\end{equation}
with relevant OPE coefficients $C_{ij{n_k}}$. We will project out the desired individual conformal block using the monodromy method. It will be briefly described how this method works but for more detailed reviews we refer to \cite{Hartman:2013mia,Fitzpatrick:2014vua,Hijano:2015rla,Anous:2016kss, Hulik:2016ifr,Banerjee:2024qgg} for details on null-vector decoupling the reader is referred to the original work \cite{BELAVIN1984333}.

The idea is to add a degenerate primary field $\hat{\psi}(z)$ to the correlator. This operator corresponds to a primary state with a null-vector at level two. 
\begin{equation}
    \tilde{\mathcal{G}}(x_j,z_i;z) = \langle \hat{\psi}(z) \mathcal{O}_H(x_1)...\mathcal{O}_H(x_N)\mathcal{O}_L(z_1)...\mathcal{O}_L(z_n)\rangle = \psi(z)\tilde{\mathcal{G}}(x_j,z_i),
\end{equation}
here the contribution from the light spectator field is assumed to decouple from the rest of the correlator. Demanding that this operator creates a state with a null-vector at level 2 leads to the following consistency condition for the function $\psi(z)$
\begin{equation}
    \psi''(z) + \frac{6}{c}T(z) \psi(z) = 0,
\end{equation}
where $T(z)$ is defined in \eqref{generalstresstensor}. This is a linear differential equation with $n+N$ regular singular points. The asymptotic behavior of the solution in the limit around one of the regular singular points can be determined, in this case the stress tensor is dominated by the single second-order pole. It is found that the two solutions approach
\begin{equation}
    \psi_{\pm}(z\rightarrow x_j) = (z-x_j)^{-\frac{1}{2}\pm \frac{1}{2}\sqrt{1-24H/c}}.
\label{localsolutions}
\end{equation}
In particular this indicates that $\psi(z)$ is a multi-valued function that picks up the monodromy
\begin{equation}
    \begin{pmatrix}
        \psi_+(z)\\
        \psi_-(z)
    \end{pmatrix}
    \rightarrow
    \begin{pmatrix}
        -e^{i\pi\sqrt{1-24H/c}} & 0\\
        0& -e^{-i\pi\sqrt{1-24H/c}}
    \end{pmatrix}
    \begin{pmatrix}
        \psi_+(z)\\
        \psi_-(z)
    \end{pmatrix}
\label{singlepointmonodromy}
\end{equation}
The following strategy allows one to project out a single conformal block from the sum \eqref{blockexpansion}, one solves the ODE for unspecified residues for the first-order poles. Subsequently one demands that around any OPE pair the monodromy matrix of the solutions is conjugate to the matrix in \eqref{singlepointmonodromy} where $H$ is now swapped out for the exchange primary scaling dimension in the OPE channel, i.e. $H\rightarrow h_p$. This procedure projects out the contribution to the OPE expansion corresponding to the intermediate primaries $\{h_p\}$. From \eqref{conformalblock} one can see that this procedure fixes the residues of the single order poles to
\begin{equation}
T(z) = \sum_{j=1}^{N} \frac{H_j}{(z-x_j)^2} - \frac{c}{6}\frac{\partial_{x_j}f_0}{z-x_j} + \frac{\partial_{x_j}f_1}{z-x_j} + \sum_{i=1}^n \frac{h_i}{(z-z_i)^2} + \frac{\partial_{z_i}f_0}{z-z_i}
\label{fullstresstensor}
\end{equation}
One now uses the demand that the monodromy of the solutions is fixed around the OPE as constraint equations to solve for the residues. Once the residues have been determined this way, integrating them yields the leading and subleading parts of the conformal block.

\subsection{Heavy-Light perturbation theory}
Once the stress tensor expectation value is of the form \eqref{fullstresstensor} it has a natural hierarchy. There are a number of `heavy' terms of order $\mathcal{O}(c)$ and set of `light' terms of order $\mathcal{O}(c^0)$, we will decompose $T(z)$ as
\begin{equation}
    T(z) = T_H(z) + T_L(z),
\end{equation}
where
\begin{equation}
    T_H(z) = \frac{\langle V|T(z)|V\rangle}{\langle V|V\rangle} = \sum_{j=1}^{N} \frac{H_j}{(z-x_j)^2} +\frac{C_j^{(0)}}{z-x_j}, 
\end{equation}
and
\begin{equation}
    T_L(z)  =   \sum_{i=1}^n \frac{h_i}{(z-z_i)^2} + \frac{c_i}{z-z_i} + \sum_{j=1}^{N}  \frac{C_j^{(1)}}{z-x_j}.
\end{equation}
where the accessory parameters have been defined through
\begin{equation}
    c_i = \frac{\partial f_1}{\partial z_i}, \;\;\; C_j^{(0)}= -\frac{c}{6}\frac{\partial f_0}{\partial x_j}, \;\;\; C_j^{(1)} = \frac{\partial f_1}{\partial x_j} 
\end{equation}
The two independent solutions $\psi_{\pm}(z)$ of the full ODE 
\begin{equation}
    \psi_{\pm}''(z) + \frac{6}{c}T(z)\psi_{\pm}(z)=0
\label{FullFuchsian}
\end{equation}
can be decomposed into
\begin{equation}
    \psi_{\pm}(z) = \rho_{\pm}(z) + \eta_{\pm}(z) + \text{subl.}
\end{equation}
Where $\rho_{\pm}(z)$ are the leading part of the solutions and $\eta_{\pm}(z)$ the subleading parts. They respectively solve the equations
\begin{equation}
    \rho_{\pm}(z)+\frac{6}{c}T_H(z)\rho_{\pm}(z) = 0,
\label{leadingFuchsian}
\end{equation}
and
\begin{equation}
    \eta_{\pm}''(z)  + \frac{6}{c}T_H(z)\eta(z) = -\frac{6}{c}T_{L}(z)\rho_{\pm(z)}.
\label{inhomogeneousFuchsian}
\end{equation}
Note that the subleading correction satisfies a inhomogeneous version of the leading equation. One of the key points of the present discussion is that the analytic structure of the first subleading correction is completely fixed by the dominant term. To demonstrate this assume a leading base of solutions $\rho_{\pm}(z)$ is given. Because there is no first-order derivative term in \eqref{leadingFuchsian} the Wronskian determinant of this basis is a constant
\begin{equation}
W(z) = \rho_{+}(z)\rho_{-}'(z) - \rho_{-}(z)\rho_{+}'(z) = \text{const.}
\end{equation}
this can easily be verified by checking that the derivative vanishes. Therefore, without loss of generality, we can normalize our solutions such that the Wronskian determinant is unity 
\begin{equation}
\rho_{+}(z)\rho_{-}'(z) - \rho_{-}(z)\rho_{+}'(z) = 1.
\label{WronskianConstraint}
\end{equation}
As mentioned before, the first subleading correction is fixed by the leading order. Our first manifestation of this phenomena is that variation of parameters \cite{Fitzpatrick:2014vua} can be used to solve \eqref{inhomogeneousFuchsian} and express the subleading correction $\eta_{\pm}(z)$ in terms of the leading solutions
\begin{equation}
    \eta_{\pm}(z) = A_{\pm}(z)\rho_{+}(z) + B_{\pm}(z)\rho_-(z)
\label{variationofparameters}
\end{equation}
Where the derivatives of the functions $A_{\pm}(z)$ and $B_{\pm}(z)$ are given by
\begin{align}
& A_{\pm}'(z) = - T_{L}(z)\rho_{\pm}(z)\rho_{+}(z),\\
& B_{\pm}'(z) =  T_{L}(z)\rho_{\pm}(z)\rho_{-}(z).
\end{align}
By integrating these expressions the subleading corrections can in principle be constructed in terms of the leading order solutions.

\subsection{A monodromy cycle for the identity block}
As mentioned in section \ref{Monodromy method} one can put constraints on the accessory parameters, i.e. the derivatives of the semi-classical conformal block, by demanding that the solutions of \eqref{FullFuchsian} satisfy predetermined monodromy conditions for loops around an OPE pair in the block's OPE channel. 

The particular block under consideration will be the identity block. The pair of identical light operators that are OPE contracted to the identity operator will be considered. Take $C$ to be a contour that encircles the two contracted light operators and no other operators. Demanding that the two light operators contract to the identity operator corresponds to demanding that the solutions of \eqref{FullFuchsian} are single-valued along the contour $C$. 

An argument will be applied similar to one utilized in \cite{Chen:2016kyz}, the leading basis of solutions $\rho_{\pm}(z)$ have branch points at the poles of the $T_H(z)$ but are analytic everywhere else. In particular this implies that the leading basis solutions $\rho_{\pm}(z)$ are manifestly single-valued along and inside the contour $C$.

The subleading correction takes the form \eqref{variationofparameters}, since the leading basis solutions $\rho_{\pm}(z)$ are single-valued the only potential source for multi-valuedness of the subleading correction is given by the coefficients $A_{\pm}(z)$ and $B_{\pm}(z)$. The monodromies picked up by the coefficients after traversing the cycle $C$ will be generically indicated by
\begin{align}
& A_{\pm}(z) \rightarrow A_{\pm}(z) + \text{Disc}_{A_{\pm}}(C), \\
& B_{\pm}(z) \rightarrow B_{\pm}(z) + \text{Disc}_{B_{\pm}}(C), 
\end{align}
where the discontinuities are formally given by
\begin{align}
& \text{Disc}_{A_{\pm}}(C) = \int_C dz \;  A_{\pm}'(z),\\
& \text{Disc}_{B_{\pm}}(C) = \int_C dz \;  B_{\pm}'(z).
\end{align}
As a result the transformation rule for the full corrected solutions can be written as
\begin{equation}
    \psi_{\pm}(z) \rightarrow \rho_{\pm}(z) + (A_{\pm}(z) + \text{Disc}_{A_{\pm}}(C))\rho_{+} + (B_{\pm}(z) + \text{Disc}_{B_{\pm}}(C))\rho_{-}
\end{equation}
The monodromy matrix, including the first subleading correction, is therefore given by
\begin{equation}
\begin{pmatrix}
\phi_+(z) \\
\phi_-(z)
\end{pmatrix}
\rightarrow
\begin{pmatrix}
1+ \text{Disc}_{A_{+}}(C)) & \text{Disc}_{B_{+}}(C)) \\
\text{Disc}_{A_{-}}(C)) & 1+ \text{Disc}_{B_{-}}(C))
\end{pmatrix}
\begin{pmatrix}
\phi_+(z) \\
\phi_-(z)
\end{pmatrix}
\end{equation}
Since the derivatives of the coefficients are given in \eqref{variationofparameters} and the leading solutions are analytic within the contour $C$ the discontinuities can be computed explicitly from Cauchy's theorem
\begin{align}
&\text{Disc}_{A_\pm}(C) = -2\pi i [h\rho_{\pm}'(z_1)\rho_{+}(z_1) + h\rho_{\pm}(z_1)\rho_{+}'(z_1) + c_1\rho_{\pm}(z_1)\rho_+(z_1)  \nonumber \\
& + h\rho_{\pm}'(z_2)\rho_{+}(z_2) + h\rho_{\pm}(z_2)\rho_{+}'(z_2) +c_2\rho_{\pm}(z_2)\rho_+(z_2)],\\ \nonumber\\
&\text{Disc}_{B_\pm}(C) = 2\pi i [h\rho_{\pm}'(z_1)\rho_{-}(z_1) + h\rho_{\pm}(z_1)\rho_{-}'(z_1) + c_1\rho_{\pm}(z_1)\rho_-(z_1)  \nonumber \\
& + h\rho_{\pm}'(z_2)\rho_{-}(z_2) + h\rho_{\pm}(z_2)\rho_{-}'(z_2) +c_2\rho_{\pm}(z_2)\rho_-(z_2)].
\end{align}
Note that $\text{Disc}_{A_-}(C) = - \text{Disc}_{B_{+}}(C)$. In order for the subleading correction to maintain the constraint that the full basis of solutions are single-valued around the cycle $C$ it is required that all four discontinuities vanish simultaneously.

\subsection{Solving the monodromy constraints}
\begin{figure}
	\centering
		\includegraphics[scale=1]{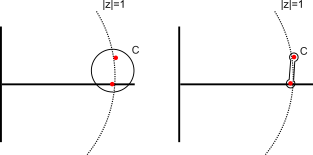}
	\caption{Deforming the contour around the two light operators to the dumbbell contour.}
	\label{dumbbellcontour}
\end{figure}

Note the following, the contour $C$ can be deformed to the limiting contour displayed in figure \ref{dumbbellcontour} around the light poles at $z_1$ and $z_2$. We can get arbitrarily close to the light operator such that the full solution approaches\footnote{This is a double limit since the light operators were assumed to form a perturbative correction to the stress tensor. For this argument to hold one needs to consider a limiting cycle around a light pole such that light pole dominates the stress tensor. Strictly speaking this requires the limit cycle to shrink faster than that $c$ goes to infinity. This point will not be considered it with more detail.} the power law \eqref{localsolutions} (where the replacement $H_j\rightarrow h_i$ has been made). The monodromy matrix around the point $z_1$ will be denoted by $\mathcal{M}$, the monodromy around the point $z_2$ is identical and therefore also given $\mathcal{M}$. The basis of solutions that has diagonal monnodromy around $z_1$ is generally not the same as the basis that diagonalizes the monodromy around $z_2$. Take the basis transformation matrix that relates these two bases to be $\mathcal{T}$. In this case the full monodromy matrix $\mathcal{M}_C$ around the contour $C$ takes the form
\begin{equation}
    \mathcal{M}_C = \mathcal{M}\mathcal{T}\mathcal{M}\mathcal{T}^{-1}.
\end{equation}
From this one concludes
\begin{equation}
    \det(\mathcal{M}_C) = \det(\mathcal{M})^2 = 1,
\end{equation}
where the last equality can be read off from \eqref{singlepointmonodromy}.

The constraints can be organized into two categories, the diagonal constraints and the off-diagonal constraint. If the diagonal constraints vanish than the off-diagonal constraint has to vanish automatically as well in order to maintain the unit determinant. The strategy will be constrain the accessory parameters in such a way that $\text{Disc}_{A_+}(C)$ and $\text{Disc}_{B_-}(C)$ vanish. Afterwards the result for the accessory parameters will be plugged into $\text{Disc}_{A_-}(C) = - \text{Disc}_{B_{+}}(C)$ to confirm that these constraints indeed also vanish as is demanded by consistency.

Note that the constraint $\text{Disc}_{A_+}(C)=0$ is given by 
\begin{equation}
    2h\rho_{+}'(z_1)\rho_{+}(z_1)  + c_1\rho_+(z_1)^2
 + 2h\rho_{+}'(z_2)\rho_{+}(z_2)  +c_2\rho_{+}(z_2)^2 = 0,
\label{firstconstraint}
\end{equation}
and note that the constraint $\text{Disc}_{B_-}(C)=0$ takes the same general form but under the substitution $\rho_+ \leftrightarrow \rho_-$
\begin{equation}
    2h\rho_{-}'(z_1)\rho_{+}(z_1)  + c_1\rho_-(z_1)^2
 + 2h\rho_{-}'(z_2)\rho_{+}(z_2)  +c_2\rho_{-}(z_2)^2 = 0.
\label{secondconstraint}
\end{equation}
Briefly taking a larger scope, the constraint that both basis solutions have to be single-valued around the contour $C$ is equivalent to the statement that the general solution to \eqref{FullFuchsian} has to be single-valued around $C$. As a consequence it is reasonable to make the Ansatz that the solutions for the accessory parameters $c_1, c_2$ are independent of the choice of particular solution of \eqref{leadingFuchsian}. This suggests the following strategy, the constraint \eqref{firstconstraint} will be promoted to the more general form $\rho_{+}(z)\rightarrow \alpha \rho_{+}(z) + \beta\rho_-(z)$, reminding ourselves that $c_i = \frac{\partial f_1(z_1,z_2)}{\partial z_i}$ leads to
\begin{align}
    &2h (\alpha \rho'_+(z_1) + \beta\rho'_-(z_1))((\alpha \rho_+(z_1) + \beta\rho_-(z_1)) \nonumber \\
    & +\frac{\partial f_1(z_1,z_2)}{\partial z_1}((\alpha \rho_+(z_1) + \beta\rho_-(z_1))^2 \nonumber \\
    & + z_1 \leftrightarrow z_2 = 0.
\label{fullconstraint}
\end{align}
notice that this contains both constraints \eqref{firstconstraint} and \eqref{secondconstraint} as special cases, subsequently selecting the specific solution of the accessory parameters to this general constraint such that the dependence on the parameters $\alpha$ and $\beta$ cancels out.

The constraint \eqref{fullconstraint} can be recognized as a partial differential equation for the light sector $f_1(z_j,x_i)$ of the semi-classical block (where in the notation of the constraint we have suppressed the dependence on the heavy sector). While the constraint equation \eqref{fullconstraint} looks complicated, it is a PDE of the generic form
\begin{equation}
    p'(z_1) + p'(z_2) +p(z_1)\frac{\partial q(z_1,z_2)}{\partial z_1} + p(z_2)\frac{\partial q(z_1,z_2)}{\partial z_2} = 0.
\end{equation}
Which has a general solution given by
\begin{equation}
    q(z_1,z_2) = -\log(p(z_1))  -\log(p(z_2)) + G\left(\int_{z_1}^{z_2} \frac{dz'}{p(z')}\right),
\end{equation}
where $G(x)$ is an undetermined differentiable function. As a consequence the solution \eqref{fullconstraint} is found to be given by
\begin{align}
    & f_1(z_1,z_2) = -2h\log \left(\alpha \rho_+(z_1) + \beta\rho_-(z_1)\right) \nonumber \\
    &-2h\log \left(\alpha \rho_+(z_2) + \beta\rho_-(z_2)\right) \nonumber \\
    & + G\left(\int_{z_1}^{z_2} \frac{dz'}{\left(\alpha \rho_+(z') + \beta\rho_-(z')\right)^2}\right).
\label{constraintproblem}
\end{align}
Since this expression has to solve the constraint for all values of the parameters $\alpha,\beta$ simultaneously the goal is to find a function $G(x)$ that ensures that $f_1(z_1,z_2)$ is actually independent of these parameters. It will be useful to establish the following relation
\begin{align}
    &\frac{d}{dz} \frac{\rho_+(z)}{\alpha \rho_+(z) + \beta\rho_-(z)} = \frac{\beta (\rho_-(z)\rho_+'(z)-\rho_+(z)\rho_-'(z))}{(\alpha \rho_+(z)+\beta \rho_-(z))^2} \nonumber \\
    & = \frac{-\beta}{(\alpha \rho_+(z)+\beta \rho_-(z))^2},
    \label{handyderivative}
\end{align}
where the last equality is due to the Wronskian normalization condition \eqref{WronskianConstraint}. This allows for the integral in the argument of $G(x)$ to be resolved
\begin{align}
    & \int_{z_1}^{z_2} \frac{dz'}{(\alpha \rho_+(z')+\beta\rho_-(z'))^2} = -\frac{1}{\beta} \int_{z_1}^{z_2} dz' \frac{d}{dz'} \frac{\rho_+(z')}{\alpha\rho_+(z')+\beta\rho_-(z')} \nonumber \\
    & = \frac{\rho_+(z_1)\rho_-(z_2) - \rho_+(z_2)\rho_-(z_1)}{(\alpha\rho_+(z_1)+\beta\rho_-{z_1})(\alpha\rho_+(z_2)+\beta\rho_-{z_2})}.
\end{align}
Plugging this expression into \eqref{constraintproblem} it is found that the specific choice
\begin{equation}
    G(x) = -2h\log(x)
\end{equation}
causes the dependence on $\alpha$ and $\beta$ to cancel out and reduces the expression for the leading light dependence of the semi-classical block to
\begin{equation}
    \boxed{f_1(z_1,z_2) -2h \log\left(\rho_+(z_1)\rho_-(z_2) - \rho_+(z_2)\rho_-(z_1)\right),}
\label{fsolution}
\end{equation}
which is manifestly independent of $\alpha$ and $\beta$. In the next subsection it will be established that the formula \eqref{fsolution} has a natural CFT interpretation. As a direct consequence it will make it easy to check for consistency by showing that the formula \eqref{fsolution} also causes the off-diagonal constraint to vanish.

\subsection{Off-diagonal constraints and the link to known results}
The expression
\begin{equation}
    f_1(z_1,z_2) -2h \log\left(\rho_+(z_1)\rho_-(z_2) - \rho_+(z_2)\rho_-(z_1)\right)
\end{equation}
leads to the following expression for the light sector of the Virasoro identity block
\begin{equation}
    Q(z_1,z_2) \equiv e^{\frac{c}{6}f_1(z_1,z_2)} = \frac{1}{\left(\rho_+(z_1)\rho_-(z_2) - \rho_+(z_2)\rho_-(z_1)\right)^{2h}}.
\label{lightsector}
\end{equation}
Which was quoted in \eqref{mainresult} as the main result of this paper. While this expression might look unfamiliar it is in fact a generalization of a known expression \cite{Fitzpatrick:2015zha} and equivalent to an expression utilized in \cite{Banerjee:2018tut}. Recall that the Wronskian condition
\begin{equation}
\rho_{+}(z)\rho_{-}'(z) - \rho_{-}(z)\rho_{+}'(z)=1,
\end{equation}
can be inverted \cite{Hulik:2016ifr} in terms of a single function $u(z)$
\begin{equation}
    \rho_{+}(z) = \frac{1}{\sqrt{u'(z)}}, \;\;\;\;\;\; \rho_-(z) = \frac{u(z)}{\sqrt{u'(z)}}.
\label{Wronskianinversion}
\end{equation}
It can easily be checked that as a consequence of
\begin{equation}
    \rho_{\pm}''(z) + \frac{6}{c}T_H(z)\rho_{\pm}(z) =0,
\end{equation}
that $u(z)$ solves the Schwarzian equation
\begin{equation}
    S[u(z),z] = \frac{12}{c}T_H(z)
\end{equation}
Substituting \eqref{Wronskianinversion} into \eqref{lightsector} leads to
\begin{equation}
    Q(z_1,z_2) = u'(z_1)^{h}u'(z_2)^{h}\frac{1}{(u(z_1)-u(z_2))^{2h}}.
\label{twopointfunction}
\end{equation}
This can be recognized as the conformal two-point function after a conformal transformation $z\rightarrow u(z)$, where $u(z)$ solves the Schwarzian equation. As promised, the reformulation of the solutions in terms of the parametrization \eqref{Wronskianinversion} makes is particularly easy to check that the off-diagonal constraint vanishes as well. The off-diagonal constraint $\text{Disc}_{A_-}(C) = -\text{Disc}_{B_+}(C)=0$ is proportional to
\begin{align}
    & \text{Disc}_{A_-}(C)\propto \sum_{i=1}^2 h\partial_{z_i} \rho_-(z_i)\rho_+(z_i) + c_i \rho_+(z_i)\rho_-(z_i)\nonumber \\
    & = 2h + \frac{u(z_1)(c_1 u'(z_1)-hu''(z_1))}{u'(z_1)^2} + \frac{u(z_2)(c_2 u'(z_2)-hu''(z_2))}{u'(z_2)^2}.
\label{uconstraint}
\end{align}
The accessory parameters in turn are given by
\begin{align}
    c_i = \frac{\partial f_1}{\partial z_i} = -2h \partial_{z_i} \log\left(\frac{u(z_2)-u(z_1)}{\sqrt{u'(z_1)u'(z_2)}}\right)
\label{uaccessoryparameters}
\end{align}
Inserting \eqref{uaccessoryparameters} into \eqref{uconstraint} shows that these accessory parameters indeed annihilate the off-diagonal constraints.

In the particular case where the heavy sector consists of only two identical heavy operators located at the origin and infinity it can be found that the heavy part of the stress tensor expectation value reduces to
\begin{equation}
    T_H(z) = \frac{H}{z^2}
\end{equation}
In this case the Schwarzian equation is simply solved by a power law solution
\begin{equation}
    u(z) = z^{\sqrt{1-24H/c}},
\end{equation}
substituting this solution into \eqref{twopointfunction} reproduces the result for the identity block presented in \cite{Fitzpatrick:2015zha}.

\section{Analytic continuation to Lorentzian time}
\label{Lorentziantime}
\begin{figure}
	\centering
		\includegraphics[scale=1]{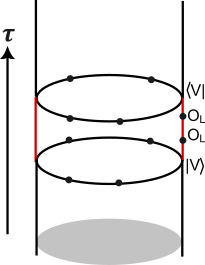}
	\caption{Visual representation of the time-ordering of the Euclidean correlator $\langle V|O_LO_L|V\rangle$ on the Euclidean cylinder.}
	\label{EuclideanTimeOrdering}
\end{figure}

The object of this paper is to compute the time-dependent expectation value
\begin{equation}
   \Phi(t_1,t_2) =  \frac{\langle V|O_L(t_2)O_L(t_1)|V\rangle}{\langle V|V\rangle}.
\end{equation}
Here the heavy state is created by acting with a finite though generic number of heavy operators on the vacuum state
\begin{equation}
    |V\rangle = O_H(x_1)...O_H(x_n)|0\rangle.
\end{equation}
As a consequence the state $|V\rangle$ is time-dependent. This means the expectation value $\Phi(t_1,t_2)$ has a non-trivial causal structure that needs to be considered. The goal is to obtain the Lorentzian causal structure through means of analytic continuation from the relevant Euclidean correlator. In a theory where the spectrum of the Hamiltonian is unbounded from above the only Euclidean time-ordering that is well defined is the one that coincides with the operator ordering. Forgetting about the Lorentzian time-labels for a moment, consider a Euclidean correlator \textit{on the cylinder} of the operators $O_L(\zeta_i,\bar{\zeta}_i)$ and $O_H(\xi_j,\bar{\xi}_j)$ with Euclidean coordinates $\zeta_i = \tau_i +i\phi_i,\; \bar{\zeta}_i = \tau_i - i\phi_i$ and $\xi_j = \tau_j + i\phi_j, \; \bar{\xi}_j = \tau_j - i\phi_j$. Assume that the Euclidean time labels $\tau_i, \tau_j$ respect a time-ordering that matches the operator ordering, see figure \ref{EuclideanTimeOrdering}

\begin{figure}
	\centering
		\includegraphics[scale=0.8]{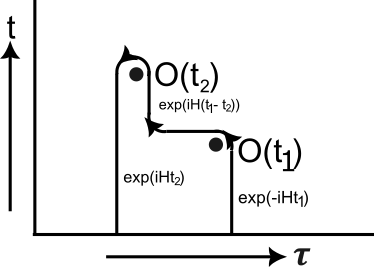}
	\caption{The Lorentzian time evolution of the operators $O(t_1)$ and $O(t_2)$ that takes the Eucldian correlator to the Lorentzian expectation value $\Phi(t_1,t_2)$.}
	\label{SchwingerKeldysh}
\end{figure}

We can now consider the analytic continuation to Lorentzian time from the Heisenberg picture perspective. Setting the initial time of the state to $t_0 = 0$ gives us the following picture
\begin{equation}
  \Phi(t_1,t_2) = \langle V|e^{iHt_2}Oe^{-iH(t_2-t_1)}Oe^{-iHt_1}|V\rangle
\end{equation}
This leads to a Schwinger-Keldysh-like contour as shown in figure \ref{SchwingerKeldysh}. Borrowing the language of \cite{Kundu:2025jsm}, this evolution can be considered from the perspective of a Minkowski spacetime diagram from which it can be seen that the compactness of the spatial direction introduces some subtleties.

\begin{figure}
	\centering
	\begin{subfigure}{.4\textwidth}	
        \centering
        \includegraphics[scale=1]{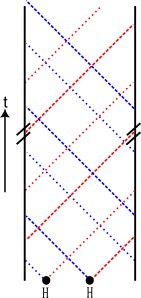}
    \end{subfigure}
    \begin{subfigure}{.4\textwidth}
        \centering
		\includegraphics[scale=1]{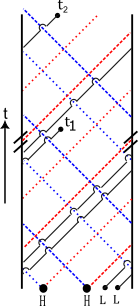}
    \end{subfigure}
    \caption{\textbf{Left:} The Lorentzian cylinder is obtained by identifying the two boundaries. The lightcone structure of the initial state is represented by the dashed lines. The red dashed lines represent the rightmoving lightcones and the blue dashed lines the leftmoving lightcones. \textbf{Right:} Turning on Lorentzian time-evolution by analytically continuing $z$ while keeping $\bar{z}$ fixed results in the following trajectories on the Minkowski diagram. Every time the light operator crosses a lightcone of the heavy sector results in a half-monodromy picked up by continuing the Euclidean correlator across a branch cut.}
    \label{lightcones}
\end{figure}

Consider for purposes of demonstration the simplest non-trivial example of a state $|V\rangle$ on that is created by acting with only two heavy operators
\begin{equation}
    |V_2\rangle = O_H(\xi_1,\bar{\xi}_1)O_H(\xi_2,\bar{\xi}_2)|0\rangle.
\end{equation}
The expectation $\langle V_2|O_LO_L|V_2\rangle$ is sensitive to the causal structure of the state. The lightcone structure of this initial state is represented in the left diagram of figure \ref{lightcones}

We will evolve both light operators initially simultaneously by dialing $\zeta_1, \zeta_2$ while keeping $\bar{\zeta}_1, \bar{\zeta}_2$ fixed, at some point we stop the evolution of $\zeta_1$ and keep going with $\zeta_2$ for a while longer. This results in the rightmoving shifts displayed in the right diagram of figure \ref{lightcones} with a final time separation between $t_1$ and $t_2$. Turning on the time-evolution of the light operators, as displayed in the right diagram of figure \ref{lightcones} results in a set of branch discontinuities, every time a light operator crosses the lightcone of a heavy operator the Euclidean correlator picks up a half-monodromy from the corresponding branch discontinuity \cite{Kundu:2025jsm}.

\subsection{Mapping to the radial plane}
\begin{figure}
	\centering
		\includegraphics[scale=1]{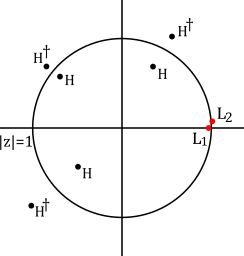}
	\caption{Visual representation of the time-ordering of the holomorphic sector of the Euclidean correlator $\Phi_E(z_i,\bar{z}_i)$ on the Euclidean cylinder. The light operators have been chosen to be inserted (close to) the unit circle $|z|=1$. To maintain the correct radial ordering the ket operators are located inside the unit circle and the adjoint bra-operators are reflected around the unit circle with respect to their ket-operator.}
	\label{radialordering}
\end{figure}

The strategy outlined in the previous section depends upon the analytic continuation of a Euclidean correlator. The relevant time-ordering structure was briefly discussed on the Euclidean cylinder, but chiefly we are concerned with computing the Euclidean correlator in conventional radial quantization. The two coordinate systems are simply related through
\begin{align}
    & z = e^{\zeta} = e^{\tau+i\phi}, \\
    & \bar{z} = e^{\bar{\zeta}}=e^{\tau-i\phi}.
\end{align}
On the radial plane a Hermitian operator with scaling dimension $h$ is related to its adjoint through the relationship 
\begin{equation}
    O(z,\bar{z})^{\dagger} = \frac{1}{z^{2\bar{h}}}\frac{1}{\bar{z}^{2h}} O(1/\bar{z},1/z),
\label{conjugation}
\end{equation}
where the real surface constraint, $\bar{z}=z^*$ has been assumed for the initial Euclidean correlator. The Euclidean version of the expectation value, i.e.
\begin{equation}
    \Phi_E(z_i,\bar{z}_i)=\langle V|O_L(z_2,\bar{z}_2)O_L(z_1,\bar{z}_1)|V\rangle,
\end{equation}
therefore takes the very explicit form
\begin{align}
    & \Phi_E(z_i,\bar{z}_i) = \nonumber \\
    &\langle O_H(1/\bar{x}_n,1/x_n)...O_H(1/\bar{x}_1,1/x_1)\; O_L(z_2,\bar{z}_2)O_L(z_1,\bar{z}_1)\; O_H(x_1,\bar{x}_1)...O_H(x_n,\bar{x}_n)\rangle,
\end{align}
where the position-dependent prefactors of \eqref{conjugation} have been stripped off. Maintaining the Euclidean time-ordering of figure \ref{EuclideanTimeOrdering} on the radial plane leads to the schematic radial ordering displayed in figure \ref{radialordering}

The next step is to perform the analytic continuation to Lorentzian time. The real-valued Euclidean time coordinates $\tau_i$ will be extended to complex valued numbers
\begin{equation}
    \tau_i\rightarrow \tau_i + i t_i,
\end{equation}
that will live on the upper-half plane. On the radial plane it can be easily seen that this results in the operator labels picking up complex phases
\begin{align}
    z_i= e^{\tau_i+i\phi_i} \rightarrow e^{\tau_i + it_i + \phi_i} = e^{it_i}z_i,\\
    \bar{z}_i= e^{\tau_i-i\phi_i} \rightarrow e^{\tau_i + it_i - \phi_i} = e^{it_i}\bar{z}_i.
\end{align}
The contour displayed in the right diagram of figure \ref{lightcones} involves a trajectory where $\tilde{z} = t+\phi $ is continued while $\tilde{\bar{z}}= t - \phi$ is held fixed. As such the spatial coordinate is moved along with the Lorentzian time in order to compensate the change in $t$
\begin{equation}
    \phi_i(t) = \phi_i +t.
\end{equation}
This leads to the following trajectory
\begin{equation}
 z_i(t)= e^{2it}z_i, \;\;\;\;\;\;\bar{z}_i(t)=  \bar{z}_i.
\label{trajectoriesz}
\end{equation}
This provides a prescription to continue the Euclidean correlator to the right branch in order to obtain the Lorentzian correlator with the correct causal structure.

\subsection{Applying the conformal block construction}
To obtain the Lorentzian expectation value $\Phi(t_1,t_2)$ two ingredients are required, the appropriate Euclidean correlator and the appropriate contour to analytically continue this Euclidean correlator to the Lorentzian time configuration. In the previous subsection the appropriate contour was obtained. In the above section a formula was derived for the light operator contribution to the identity conformal block. Specifically the expression \eqref{lightsector} which took the form
\begin{equation}
    Q(z_1,z_2) = \frac{1}{\left(\rho_+(z_1)\rho_-(z_2) - \rho_+(z_2)\rho_-(z_1)\right)^{2h}},
\label{lightsector2}
\end{equation}
where the functions $\rho_{\pm}(z)$ are the Wronskian normalized solution of the differential equation
\begin{equation}
    \rho_{\pm}(z)+\frac{6}{c}T_H(z)\rho_{\pm}(z) = 0,
\end{equation}
with normalized stress energy tensor expectation value of the heavy sector
\begin{equation}
T_H(z) = \frac{\langle V|T(z)|V\rangle}{\langle V|V\rangle}.
\end{equation}
The expression \eqref{lightsector2} provides the identity block contribution to the Euclidean correlator. 

The derivation of \eqref{lightsector2} only assumed that the two light operators are contracted to the identity operator. The rest of the OPE channel was left implicit as it does not affect the overall form of expression \eqref{lightsector2}. What is affected by the choice of OPE channel is the basis of solutions $\rho_{\pm}(z)$. Each heavy operator location introduces a branch point in the functions $\rho_{\pm}(z)$, the way these branch points are paired up by the way they are connected by branch cuts selects an OPE channel. By construction \eqref{lightsector2} is independent of choice of basis, reflecting the earlier statement that the light sector is insensitive to the choice of OPE channel in the heavy sector. A basis of solutions is assumed to exist such that the branch cuts extend between an operator and its mirrored adjoint operator. 

\begin{figure}
	\centering
		\includegraphics[scale=1]{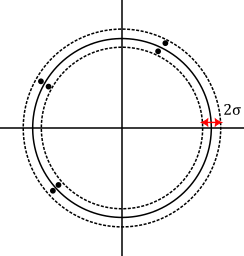}
	\caption{Restricting the heavy operator to lie within a thin annulus of width $2\sigma$ centered around the unit circle gives us a controllable parameter with which the identity block dominance of the Euclidean correlator can be controlled.}
	\label{annulus}
\end{figure}

The expression \eqref{lightsector2} only provides the identity block contribution to the full Euclidean correlator. This issue can be ameliorated by restricting the heavy operator insertions to a thin annulus of width $2\sigma$ centered around the unit circle, see figure \ref{annulus}. This provides a controllable parameter $\sigma$ that allows for control over the level of dominance of the identity block contribution to the OPE channel where the each heavy operator is contracted with its adjoint mirror image. In this case Smaller $\sigma$ corresponds to a larger parametric suppression of the higher conformal blocks.

\subsection{Time-evolution of the conformal block}
\begin{figure}
	\centering
		\includegraphics[scale=1]{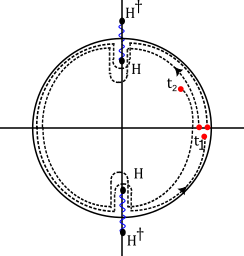}
	\caption{The time-evolution of the light operators on the radial plane that corresponds to the trajectory through Minkowski space displayed in the right diagram of figure \ref{lightcones}. Note that due to the multi-valuedness of $\rho_{\pm}(z)$ a set of half-monodromies are picked up every lightcrossing time in perfect analogy with the lightcone crossings in figure \ref{lightcones}.}
	\label{radialtimeevolution}
\end{figure}

Given the Euclidean correlator expressed in terms of the identity conformal block the analytic continuation to Lorentzian time can be performed by applying the continuation \eqref{trajectoriesz}
\begin{align}
    & z_i(t)= e^{2it}z_i, \\
    & \bar{z}_i(t) = \bar{z}_i.
\end{align}
Applying this contour on the complex plane to the conformal block \eqref{lightsector2} leads to a trajectory for the light operators on the complex plane. These trajectories are displayed in figure \ref{radialtimeevolution}

By considering the curve parametrized by \eqref{trajectoriesz} full `lightcrossing time' in the Minkowski diagram \ref{lightcones} corresponds to two full circles on the complex plane. Looking at the example of that diagram, consider a state $|V\rangle$ consisting of two heavy operators as an example. This example is displayed in figure \ref{radialtimeevolution}. Note that each time one of the light operator comes close to a pair of heavy operators the branch cut that runs between them has to be circumvented. As a result a half-monodromy is picked up. On the Minkowski diagram, each time a light operator is advanced a full lightcrossing time four half-monodromies are picked up alternating between each of the two heavy operators. In perfect correspondence, every lightcrossing time corresponds to rotating the light operator around the unit circle twice, each rotation two branch cut are circumvented one belonging to each heavy operator.


\section{Out-of-equilibrium physics, implication of Floquet's theorem}
\label{Floquetsection}

The light operators, on the Euclidean radial plane where initially decided to live on the unit circle. As discussed in the previous section, the time-evolution of the light operators is encoded in their trajectory around the unit circle\footnote{In many ways the logic of this section runs in parallel with \cite{Banerjee:2018tut}, though it refines that analysis on some points. The author owes a great debt of gratitude to S. Banerjee and J.W. Brijan and also K. Papadodimas for many discussions on the subject.}. This is suggestive that the knowledge of the light-sector on the entire complex plane is redundant to understand the late time physics of this configuration. In this section the following topic will be discussed, what remains after restricting the expression \eqref{lightsector2} to the unit circle. In the process it will become clear that a new kind of structure will emerge that is particularly well-suited for understanding the late-time limit.


\subsection{Expressing the conformal block in terms of Lorentzian time}
To clarify the procedure, first the conformal identity block \eqref{lightsector2} on the radial plane will be mapped to the Euclidean cylinder. The solutions $\rho_{\pm}$ are not invariants under such conformal transformations. Since the ODE
\begin{equation}
    \rho''(z) + \frac{6}{c}T_H(z)\rho(z)=0,
\label{Fuchsian2}
\end{equation}
contains $T_H(z)$ which transforms under conformal transformations $z\rightarrow w(z)$ the solutions will transform as well. The derivation of the transformation rule is provided in appendix \ref{ODEtransformation}. But $\rho(z)$ solves \eqref{Fuchsian2} then after the replacement $z\rightarrow w(z)$ the function $\tilde{\rho}(w)$ will solve the equation
\begin{equation}
    \frac{d^2}{dw^2}\tilde{\rho}(w) + \frac{6}{c}\tilde{T}(w)\tilde{\rho}(w)=0,
\end{equation}
where $\tilde{T}(w)$ is given by
\begin{equation}
\tilde{T}(w) = z'(w)^2 T(w) + \frac{c}{12}\{z,w\}
\end{equation}
and 
\begin{equation}
    \tilde{\rho}(w) = \frac{1}{\sqrt{z'(w)}}\rho(w)
\end{equation}
Note that this latter expression implies that the light sector \eqref{lightsector2} transforms as a conformal two-point function under the substitution $\rho(z) \rightarrow \tilde{\rho}(w)$
\begin{equation}
    \tilde{Q}(w_1,w_2) = \frac{1}{(\tilde{\rho}_+(w_1)\tilde{\rho}_-(w_2)-\tilde{\rho}_+(w_2)\tilde{\rho}_-(w_1))^{2h}} = \frac{z'(w_1)^h z'(w_2)^h}{(\rho_+(w_1)\rho_-(w_2)-\rho_+(w_2)\rho_-(w_1))^{2h}}
\end{equation}
The mapping from the radial plane to the Euclidean cylinder will be performed through means of the transformation rule
\begin{equation}
z=e^{2ix}
\label{planetocylinder}
\end{equation}
This neatly dovetails into the expression \eqref{trajectoriesz} for the trajectory of the analytic continuation to Lorentzian time, this justifies the replacement $x\rightarrow t$. This results in the expression for the transformed differential equation
\begin{equation}
    \tilde{\rho}''(t) + \left(1-\frac{24}{c}e^{4it}T(t)\right)\tilde{\rho}(t)=0.
\end{equation}
where $\tilde{\rho}(t)$ is related to $\rho(t)$ through
\begin{equation}
    \tilde{\rho}(t) = \frac{e^{it}}{\sqrt{2i}}\rho(t).
\end{equation}
As a result, in terms of the original basis of solutions the light sector up to irrelevant constant factors takes the form 
\begin{equation}
    Q(t_1,t_2) = \frac{e^{2iht_1} e^{2iht_2}}{(\rho_+(t_1)\rho_-(t_2)-\rho_+(t_2)\rho_-(t_1))^{2h}}.
\label{lightsector3}
\end{equation}
With an underlying assumption that $t_2>t_1$.

\subsection{Reality condition and Hill's equation}
The defining equation from which the light sector of the conformal block is derived:
\begin{equation}
    \tilde{\rho}''(t) + \left(1-\frac{24}{c}e^{4it}T(t)\right)\tilde{\rho}(t)=0,
\label{Hillseq}
\end{equation}
has a property that is not initially obvious. The potential $\mathcal{V}(t)$
\begin{equation}
\mathcal{V}(t) = 1-\frac{24}{c}e^{4it}T(t),
\label{Hillpotential}
\end{equation}
is real-valued. This follows from a reflection property of $T(z)$ combined with the adjoint relationship \eqref{conjugation} resulting in
\begin{align}
    &(z^2 T(z))^* = \left(z^2 \frac{\langle V|T(z)|V\rangle}{\langle V|V\rangle}\right)^{*} \nonumber \\
    & = (z^*)^2 \frac{\langle V|\frac{1}{(z^*)^4}T(1/z^*)|V\rangle}{\langle V|V\rangle} = \frac{1}{(z^*)^2}T(1/z^*).
\end{align}
On the unit circle $z=1/z^*$ therefore, restricted to the unit circle
\begin{equation}
(z^2 T(z))^* = z^2 T(z), \;\;\;\; \text{if}\; z\in S^1 \;\;\; \implies \;\; z^2 T(z) \in \mathbb{R} , \;\;\;\; \text{if}\; z\in S^1.
\end{equation}
From the substitution \eqref{planetocylinder}, i.e. $z=\exp(2it)$ the conclusion follows that $\mathcal{V}(t)$ is real-valued. From this it follows that the equation \eqref{Hillseq} is a second-order linear ODE with a real-valued periodic potentials. As a result it falls in the class of Hill's equations \cite{MagnusWinkler}.

\subsection{Floquet's theorem and late-time dynamics}
The generic behavior of solutions to Hill's equation are encapsulated by Floquet's theorem. Given a basis of solutions $\rho_{1,2}(t)$ to Hill's equation \eqref{Hillseq} with initial conditions
\begin{equation}
    \rho_1(0)=1, \;\;\; \rho_1'(0)=0, \;\;\; \rho_2(0) = 0, \;\;\; \rho_2'(0)=1,
\end{equation}
define the characteristic equation as
\begin{equation}
    y^2 - (\rho_1(\pi) + \rho_2'(\pi))y +1 =0.
\end{equation}
The roots of this quadratic equation are the characteristic exponents of the differential equation
\begin{equation}
    y_1 = e^{i\alpha \pi}, \;\;\; y_2 = e^{-i\alpha\pi}.
\end{equation}
(Part of) Floquet's theorem can now be stated as: if the roots $y_{1,2}$ do not coincide $y_1\neq y_2$, then there exists a basis of solutions $\rho^F_{\pm}(t)$ to \eqref{Hillseq} of the form
\begin{equation}
    \rho^F_{\pm}(t) = e^{\pm i\alpha t}p_{\pm}(t),
\label{unstableFloquet}
\end{equation}
with periodic functions $p_\pm(t+\pi)= p_\pm(t)$.

An implication of this theorem is that there are two generic cases for the solutions to \eqref{Hillseq}. Either $\alpha$ is real, in which case all solutions are bounded, or $\alpha$ is complex, in which case there is one exponentially growing mode and one exponentially decaying mode.

Consider the case where $\alpha$ is complex. Linearity and real-valuedness allows one to choose a real-valued basis, subsequently demanding that these solution stay real under a full period $\pi$ ensure that $\alpha$ is purely imaginary $\alpha=i|\alpha|$. In this case $\rho^F_+(t)$ is the decaying mode and $\rho^F_-(t)$ the growing mode. The basis of solutions $\rho_{\pm}(t)$ in \eqref{lightsector3} can be decomposed into the Floquet basis
\begin{align}
    & \rho_{+}(t) = a \rho^F_+(t) +b \rho^F_-(t), \\
    & \rho_{-}(t) = c \rho^F_+(t) +d \rho^F_-(t).
\end{align}
This provides structure for the time-evolution of the light sector \eqref{lightsector3}. By construction, the light sector is basis independent, it can easily be verified that if $ad-bc=1$ the light sector takes the form 
\begin{equation}
    Q(t_1,t_2) = \frac{e^{2ih(t_1+t_2)} }{(e^{-|\alpha| (t_1-t_2)}p_+(t_1)p_-(t_2)-e^{-|\alpha|(t_2-t_1)}p_+(t_2)p_-(t_1))^{2h}}.
\label{lightsector4}
\end{equation}
Consider the time-evolution depicted in figure \ref{lightcones} as consisting of two parts. An initial part where both light operator $O_L(t_1)$ and $O_L(t_2)$ are evolved side-by-side. And a second part where $O_L(t_1)$ is held fixed and $O_L(t_2)$ is evolved alone. During the first phase of evolution the difference $t_1-t_2$ is preserved and as a consequence $Q(t_1,t_1+\epsilon)$ is a periodic function of $t_1$ with period $\pi$, i.e. the `light-crossing time' in the Minkowski diamond. 

During the second phase of evolution $t_1$ is held fixed and $t_2$ is evolved independently, in this case there is an initial period where the decaying term in the denominator dies out. After this initial time $Q(t_1,t_2)$ takes the schematic form
\begin{equation}
    Q(t_2>t_1) \propto \frac{e^{2ih(t_1+t_2})}{p_+(t_1)^{2h}p_-(t_2)^{2h}} \times e^{2h\, |\alpha|(t_1-t_2)}.
\label{lightsector5}
\end{equation}
This expression consists of an overall periodic factor multiplied with an exponentially decaying factor. Exponential decay in Lorentzian two-point function is a hallmark of thermal physics. The holomorphic part factor of the thermal Lorentzian CFT two-point function of primary operators is at intermediate time-scales of the general form
\begin{equation}
    \text{Tr}\left( e^{-(\beta -it)H}Oe^{-iHt}O \right) \sim e^{-2\pi h Tt}
\end{equation}
As a consequence it makes sense to compare the exponent of the decaying factor of \eqref{lightsector5} with that of the thermal two-point function. This leads to a relationship between the Floquet root and the effective temperature $T_{|V\rangle}$ experienced by the light operators
\begin{equation}
    T_{|V\rangle} = \frac{|\alpha|}{\pi}.
\label{temperature}
\end{equation}
The conclusion is that in the case that Hill's equation contains a solution that it unbounded it immediately implies that the expectation value of light operators act as if they were in contact with a thermal bath.


\section{CFT thermodynamics and Virasoro coadjoint orbits}
\label{thermodynamicsandorbits}
The Floquet exponent from the last section has an explicit meaning in Virasoro representation theory as an orbit invariant of a Virasoro coadjoint orbit. This suggests another direct link between Virasoro representation theory and CFT thermodynamics that leads to a small puzzle. The result of the previous sections suggest that the effective physics of the light-operators is controlled by a single number, the Floquet exponent. Despite that a priori the state $|V\rangle$ from a CFT perspective is quite complicated. In particularly, generically it would be expected that $|V\rangle$ is a complicated superposition of many CFT states from different conformal representations. 

Instead it was found that the data is contained within a derived quantity from the stress tensor, $\mathcal{V}(t)$ which has the structure of (part of) a Virasoro coadjoint vector. The Floquet exponent, in turn, is an orbit invariant of a Virasoro coadjoint orbit, the set of all Virasoro coadjoint vectors that are connected by monotonic reparametrizations of the variable $t$. The Virasoro coadjoint orbits, at least the ones that will be considered here, are the classical limits of Virasoro representations \cite{Alekseev:1988ce,Alekseev:1990mp}. They form disconnected manifolds. As such, a Virasoro coadjoint vector can only be contained in one orbit at a time. A statement that is add odds with the claim that $|V\rangle$ is superposition of many different Virasoro representations.

This section will start with a derivation that contains all the physics of the claim above without any required knowledge about Virasoro coadjoint orbits. This argument is derived from a discussion in \cite{Balog:1997zz}. Following this derivation there will be an argument based on the disconnectedness of different orbits. A very concise introduction to Virasoro coadjoint orbits will be provided in appendix \ref{vcoadorbits}.

The starting point of the argument is Hill's equation \eqref{Hillseq}
\begin{equation}
    \bar{\rho}''(t) + \mathcal{V}(t)\bar{\rho}(t)=0,
\end{equation}
where the potential $\mathcal{V}(t)$ is given by 
\begin{equation}
\mathcal{V}(t) = 1-\frac{24}{c}e^{4it}T(t).
\end{equation}
As established in appendix \ref{ODEtransformation}, under an orientation-preserving diffeomorphism of the circle $t\rightarrow s(t)$ the ODE transforms to
\begin{equation}
    \tilde{\rho}''(s) + \tilde{\mathcal{V}}(s)\tilde{\rho}(s)=0.
\end{equation}
The objects in this transformed equation are respectively given by
\begin{align}
    & \tilde{\mathcal{V}}(s) = t'(s)^2\mathcal{V}(t(s)) + \frac{1}{2}\{t,s\}, \label{Hilltransformation} \\
    & \tilde{\rho}(s) = \frac{1}{\sqrt{t'(s)}}\rho(t(s),
\end{align}
where the Schwarzian derivative is given by
\begin{equation}
    \{t(s),s\} = \frac{t'''(s)}{t'(s)} -\frac{3}{2}\left(\frac{t''(s)}{t'(s)}\right)^2.
\end{equation}
The Floquet basis of solutions can itself be used to construct a special kind of conformal transformation. The working assumption is still that $\alpha$ is purely imaginary, i.e.
\begin{equation}
    \rho^F_{\pm}(t) = e^{\mp|\alpha|t}p_{\pm}(t).
\end{equation}
Observe that due to the form of $Q(t_1,t_2)$ neither of the two solutions can ever have any zeroes. If they did the light sector \eqref{lightsector5} would have unphysical contact singularities outside of the operator insertions. Knowing that, define $u(t)$ as the ratio of solutions
\begin{equation}
    u(t)=\frac{\rho_-(t)}{\rho_+(t)},
\end{equation}
due to the Wronskian normalization condition $u(t)$ is a monotonically increasing function
\begin{equation}
u'(t) = \frac{1}{\rho_-(t)^2}>0. 
\end{equation}
and since $\rho_{\pm}(t)$ does not have any zeroes the following diffeomorphism of the circle can be constructed
\begin{equation}
    s(t) = \frac{1}{2|\alpha|}\log(u(t)),
\end{equation}
it can easily be checked that this expression satisfies $s(t+\pi)=s(t)+\pi$ and $s'(t)>0$ hence this is a diffeomorphism of the circle. Assuming $\rho_+\rho'_--\rho_-\rho'_+=1$ the Schwarzian derivative of $s(t)$ reduces to
\begin{equation}
    \{s(t),t\} = 2|\alpha|^2 s'(t)^2 -2\frac{\rho_-''}{\rho_-}.
\end{equation}
Since $\rho_-'' + \mathcal{V}\rho_-=0$ rearranging these terms gives
\begin{equation}
    \mathcal{V}(t) = -|\alpha|^2 s'(t)^2 + \frac{1}{2}\{s(t),t\}.
\label{reconstructedpotential}
\end{equation}
From the transformation rule of Hill's equation \eqref{Hilltransformation} it can be seen that $\mathcal{V}(t)$ is related by diffeomorphism to a constant potential given by $-|\alpha|^2$.

The constant Hill potential can be related back to the CFT stress tensor by means of \eqref{Hillpotential} and \eqref{planetocylinder}, these two expression combined lead to
\begin{equation}
    V(t) = -|\alpha|^2 \;\; \rightarrow \;\; T_{\alpha}(z) = \frac{c}{24}\frac{1+|\alpha|^2}{z^2}.
\end{equation}
This can be recognized as the normalized stress tensor expectation value of a primary state with a particular scaling dimension.
\begin{equation}
    T_{\alpha}(z) = \frac{\langle O_{H_{\text{eff.}}}(0)T(z)O_{H_{\text{eff.}}}(\infty)\rangle}{\langle O_{H_{\text{eff.}}}(0)O_{H_{\text{eff.}}}(\infty)\rangle}, \;\;\;\; H_{\text{eff.}}=\frac{c}{24}(1+|\alpha|^2)
\label{effectivestresstensor}
\end{equation}

There exists, at least naively, a relationship between diffeomorphism of the circle and conformal transformations of the complex plane. If $s(t)$ is a diffeomorphism of the circle, i.e. $s'(t)>0$ and $s(t+\pi)=s(t)+\pi$ then $e^{2is(t)}$ is periodic function on the interval $[0,\pi]$ with a discrete Fourier transformation
\begin{equation}
    e^{2is(t)} = \frac{1}{2\pi} \sum_{n=-\infty}^{n=\infty} c_n e^{2int}.
\end{equation}
this can be extended to the complex plane by the the substitution $z=e^{2\pi i t}$, after which the Fourier series turns into a Laurent series for a conformal transformation. 

As a result the conclusion can be drawn that if the solution to Hill's equation has an exponentially growing mode and the solutions have no zeroes that the stress-tensor $T(z)$ is conformally related to a stress tensor expectation value of a time-independent primary state\footnote{Note that the scaling dimension is bounded from below by $c/24$, therefore in AdS$_3$/CFT$_2$ the dual of such a state would be a BTZ geometry above the BTZ mass gap \cite{Banados:1992wn}} with scaling dimension $H_{\text{eff.}}= c(1+|\alpha|^2)/24$. 

This raises an interesting paradox, the computation started with light operators on an initial state $|V\rangle$ which was a fairly generic time-dependent state. As a result one would a priori expect it to be a linear superposition of states of many conformal representations. Instead, the perspective of the light operators is that they exclusively sense a stress-tensor expectation value. One that is conformally equivalent to that of a single time-independent effective primary state.


\subsection{Relationship in terms of Virasoro coadjoint orbits}
\label{relationshiptoorbits}
The argument above can be phrased separately in terms of Virasoro representation theory. All the essential physics is already contained in the previous subsection, but it can be placed on more formal footing by phrasing it in the language of Virasoro coadjoint orbits.

As reviewed in appendix \ref{vcoadorbits} a coadjoint vector of the Virasoro algebra is a doublet consisting of a quadratic differential and a real number
\begin{equation}
(\mathcal{V}(t)dt^2,b) \in \overline{\text{diff}}^*(S^1).
\end{equation}
The coadjoint action of an adjoint vector $(\epsilon(t)\frac{d}{dt},a)\in \overline{\text{diff}}(S^1)$ on this co-vector is given by
\begin{equation}
    \text{ad}^*_{(\epsilon(t)d/dt, a)} (\mathcal{V}dt^2,b) = \left(2\epsilon'\mathcal{V}+\mathcal{V}'\epsilon +\frac{b}{24\pi}\epsilon''', 0\right)
\end{equation}
Recall that under the replacement $t\rightarrow s(t)$ the ingredients of Hill's equation
\begin{equation}
    \tilde{\rho}''(s) + \tilde{\mathcal{V}}(s)\tilde{\rho}(s)=0,
\end{equation}
transform as
\begin{align}
    & \tilde{\mathcal{V}}(s) = t'(s)^2\mathcal{V}(t(s)) + \frac{1}{2}\{t,s\}, \label{Hilltransformation2} \\
    & \tilde{\rho}(s) = \frac{1}{\sqrt{t'(s)}}\rho(t(s).
\end{align}
Therefore under infinitessimal transformations $t\rightarrow t + \epsilon(t)$, the Hill potential transforms as
\begin{equation}
\mathcal{V}(t) \rightarrow \mathcal{V}(t) - 2\epsilon'\mathcal{V} - \mathcal{V}\epsilon' -\frac{1}{2}\epsilon'''
\end{equation}
Therefore we can identify $\mathcal{V}(t)$ as part of a Virasoro coadjoint vector\footnote{The sign discrepancy in the transformation rule is due to a difference in convention between transformations and inverse transformations.}
\begin{equation}
    (\mathcal{V}(t)dt^2, 12\pi).
\end{equation}
The second entry appears somewhat ad hoc, this is a consequence of the relationship between the potential and the stress tensor expectation value \eqref{Hillpotential}, where the central charge was divided out. All Virasoro coadjoint orbits have been classified \cite{Segal1981,LazPan75,WittenCoadjointOrbits,Balog:1997zz}. The (implicit) properties required in order to arrive at thermal expression \eqref{lightsector5} pins $\mathcal{V}(t)$ down to a specific orbit. This orbit has been quantized \cite{Alekseev:1988ce,Alekseev:1990mp} and points to a highest weight representation of the Virasoro algebra.

As briefly reviewed in appendix \ref{vcoadorbits} a Virasoro coadjoint orbit is in part characterized by its stabilizer subalgebra, i.e. the subset of elements of $\overline{\text{diff}}(S^1)$ that leave a coadjoint vector $(\mathcal{V}(t)dt^2,12\pi)$ invariant. This is given by vector fields on $f(t)d/dt$ that solve the equation 
\begin{equation}
    2\epsilon'\mathcal{V}+\mathcal{V}'\epsilon +\frac{b}{24\pi}\epsilon'''=0.
\end{equation}
Locally the three independent solutions to this differential equation can be expressed in terms of products of the solutions to Hill's equation. In particular in the Floquet basis \eqref{unstableFloquet} 
\begin{equation}
    \epsilon_1(t) =  \left(\rho^F_+(t)\right)^2, \;\;\; \epsilon_2(t) = \rho_+^F(t)\rho_-^F(t), \;\;\; \epsilon_3(t) =  \left(\rho^F_-(t)\right)^2.
\end{equation}
Globally, the vector field has to satisfy the periodicity of the circle. Assuming purely imaginary $\alpha$ in the Floquet basis, then of these three the only solution can be uplifted to a global solution on the circle is $\epsilon_2(t)$. Hence the stabilizer algebra is one-dimensional. The implication is that $(\mathcal{V}(t)dt^2,12\pi)$ is an element of the quotient
\begin{equation}
    \overline{\text{diff}}(S^1)/\mathcal{S}^1,
\end{equation}
where $\mathcal{S}^1$ indicates the group of rigid rotations of the circle. Two remaining relevant properties include:
\begin{itemize}
    \item The Floquet basis \eqref{unstableFloquet} is unstable, i.e. $\alpha$ is purely imaginary.
    \item In order to avoid unphysical singularities in $\eqref{lightsector5}$ the Floquet basis function cannot have zeroes.
\end{itemize}
This is enough information to pin the orbit down\footnote{In the notation of \cite{Balog:1997zz} it would be indicated as $\mathcal{B}_0(|\alpha|)$}. This orbit has two additional properties, as discussed in the previous section, it can be brought to a constant representative coadjoint vector by a diffeomorphism transformation
\begin{equation}
    \mathcal{V}(t) = -|\alpha|^2 s'(t)^2 + \frac{1}{2}\{s(t),t\},
\end{equation}
where $s(t)$ is given by
\begin{equation}
    s(t) = \frac{1}{2|\alpha|}\log\left(\frac{\rho_-^F(t)}{\rho^F_+(t)}\right).
\label{quasisymmetricmap}
\end{equation}
The second property is related to the energy of the orbit. The energy of a coadjoint vector $(G(t)dt^2,b)$ can be defined by the integral
\begin{equation}
    E_G= \frac{1}{\pi}\int_0^\pi G(t)\,dt.
\end{equation}
The statement is that for any element of the orbit of $(\mathcal{V}(t)dt^2,12\pi)$ the energy is bounded from below and the lower bound is given by the energy of the constant representative $E_{\text{lowest}} = -|\alpha|^2$. As discussed in the previous section. Heuristically, this is suggestive of the orbit of $(\mathcal{V}(t)dt^2,12\pi)$ being the classical limit of Virasoro hightest weight representation with scaling dimension $H=c(1+|\alpha|^2)/24$.

\section{Conformal bootstrap}
\label{conformalbootstrap}
The above coarse-graining procedure experienced by light operators suggest that there is a kind of projection that eliminates most of the information of the initial state down to the data of a single representation. In this section it will be shown that in a specific case this projection is corroborated by known results from the conformal bootstrap \cite{ Cardy:2017qhl, Das:2017cnv, Collier:2019weq}.

\begin{figure}
	\centering
		\includegraphics[scale=0.8]{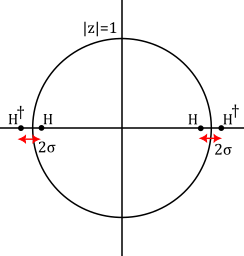}
	\caption{The distribution of four heavy operators for the expectation value $\langle V_2|O_LO_L|V_2\rangle$.}
	\label{Fourpointdistribution}
\end{figure}

For simplicity, consider the simplest non-trivial case, where the state $|V_2\rangle$ is built out of only two heavy operators
\begin{equation}
    |V_2\rangle = O(-1+\sigma)O(1-\sigma)|0\rangle.
\end{equation}
In the previous section it was assumed that the heavy operators were contained in an annulus of width $2\sigma$ (see figure \ref{annulus}). This ensured the existence of a controllable parameter $\sigma$ that tuned to what extent the identity block dominates the OPE-channel where a heavy operator is contracted with its adjoint. The resulting heavy operator distribution for the Euclidean correlator $\langle V_2|O_LO_L|V_2\rangle$ is displayed in figure \ref{Fourpointdistribution}.

\begin{figure}
	\centering
		\includegraphics[scale=0.8]{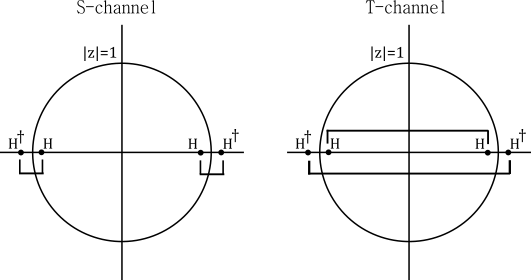}
	\caption{The S- and T-channel OPE contractions of the heavy operators in  figure \ref{Fourpointdistribution}.}
	\label{fourpointbootstrap}
\end{figure}

The conformal block decomposition will be considered in more detail. Take the mixed heavy-light conformal blocks at large $c$ as defined in section \ref{monodromymethod}
\begin{equation}
    \mathcal{F}(H,h,h_p;z_i,x_i) = e^{-\frac{c}{6}f_0(H,h_p,z_i) + f_1(H,h,h_p;x_i,z_i)+\mathcal{O}(1/c)},
\end{equation}
where $h_p$ is the exchange primary, by assumption both $f_0$ and $f_1$ only depend on order $\mathcal{O}(c^0)$ ratios of the heavy scaling dimension, e.g. $H/c$. For simplicity denote the leading term
\begin{equation}
    F(H,h_p;\sigma^2) = e^{f_0(H,h_p,x_i)},
\end{equation}
where $\sigma^2$ is the conformal cross-ratio constructed out of $x_i$. $F(H,h_p;x)$ are the conformal blocks in absence of the light operators. The
first-subleading term is denoted:
\begin{equation}
    Q(z_i,h) = e^{f_1(H,h,h_p;x_i,z_i)}.
\end{equation}
The presence of the light operators is only manifested in the subleading terms, hence the overall orthogonality of the two heavy operator contractions in not affected to leading order. Phrased differently, after contracting the two heavy OPE pairs, the only terms that survive the subsequent double sum are the diagonal terms, the two light operators fuse to the identity operator which fuses to the final identity operator that remains of the heavy sector. All other terms are suppressed. This results in the conformal block decomposition
\begin{equation}
    G(H,h;\sigma^2,z_i) \equiv \langle V_2|O_L(z_1)O_L(z_2)|V_2\rangle = \sum_{h_p} |f_{O_H O_H O_{h_p,\bar{h}_p}}|^2 Q(z_i,h,h_p)F(H,h_p;\sigma^2),
\end{equation}
where the anti-holomorphic sector has been suppressed. Since $\sigma^2$ is assumed to be a small number, the expansion above is dominated by the identity block resulting in 
\begin{equation}
    G(H,h;\sigma^2\ll 1,z_i) \approx Q(z_i,h,0)F(H,0;\sigma^2),
\end{equation}
from which we obtain what was utilized in the section \ref{Lorentziantime}
\begin{equation}
    Q(z_i,h,0) \approx \frac{\langle V_2|O_L(z_1)O_L(z_)|V_2 \rangle}{\langle V_2|V_2\rangle}.
\end{equation}
Crossing symmetry is manifested as
\begin{equation}
    G(H,h;\sigma^2,z_i) = G(H,h;1-\sigma^2 ,z_i).
\end{equation}
The left-hand side comprises the S-channel where a heavy operator is contracted with its adjoint. The right-hand side corresponds to the T-channel where instead the two in-state operators and the two out-state operators are contracted, see figure \ref{fourpointbootstrap}.

The S-channel is dominated by the identity block, this reduces the crossing equation to
\begin{equation}
    Q(z_i,h,0)F(H,0;\sigma^2) = \sum_{h_p} |f_{O_H O_H O_{h_p,\bar{h}_p}}|^2 Q(z_i,h,h_p)F(H,h_p;1-\sigma^2).
\end{equation}
An important assumption will now be added, which is that if the S-channel is dominated by the identity block the T-channel's dominant contribution comes from heavy exchange primaries ($h_p/c\sim \mathcal{O}(1)$). In the case of CFTs with an extended Cardy regime \cite{Hartman:2014oaa} the density of primary operators in this regime is very high and the sum can effectively be replaced by an integral. The anti-holomorphic factor of the heavy sector will be briefly reintroduced
\begin{align}
  &  Q(z_i,h,0)F(H,0;\sigma^2)\bar{F}(\bar{H},0,\sigma^2) \nonumber \\ 
  & = \int_0^{\infty} d\Delta_p \;  |\overline{f_{O_H O_H O_{h_p,\bar{h}_p}}}|^2 Q(z_i,h,h_p)F(H,h_p;1-\sigma^2)\bar{F}(\bar{H},\bar{h}_p ;1-\sigma^2),
\label{crosschannelintegral}
\end{align}
where $\Delta_P = h_p+\bar{h}_p$ and the barred quantity $\overline{|f|}^2$ indicates the averaged OPE coefficients squared. The conformal blocks in the T-channel can be represented by a recursion formula \cite{Zamolodchikov:1987avt} 
\begin{equation}
    F(H,h_p,z) = (16q)^{h_p-\frac{c-1}{24}}(z(1-z))^{\frac{c-1}{24} - 2H}\vartheta_3(q)^{(c-1)/2-16H}\mathcal{H}(H,h_p),
\end{equation}
here $\vartheta_3(q)$ is the Jacobi theta-function and $\mathcal{H}(H,h_p)$ is a $z$-independent function that is fixed by recursion. The modulus $q$ is related to the conformal cross-ratio $z$ through
\begin{equation}
    q = e^{i\pi\tau}, \;\;\;\; \tau = i\frac{K(1-z)}{K(z)},
\end{equation}
where $K(z)$ is the elliptic integral of the first kind
\begin{equation}
    K(z) = \frac{1}{2}\int_0^1  \frac{dt}{\sqrt{t(1-t)(z-t)}}.
\end{equation}
In the regime where $z\rightarrow 1$ the modulus approaches $q\rightarrow 1$. If $h_p$ is heavy at large central charge then the recursion factor reduces to
\begin{equation}
    \mathcal{H}(H,h_p) \overset{c\gg1}{\overset{h_p/c = \text{finite}}\rightarrow} 1+ \mathcal{O}\left(\frac{1}{c}\right).
\end{equation}
As a result the combined relevant part of the holomorphic and anti-holomorphic block reduces in the relevant regime to
\begin{equation}
    F(H,h_p,z)\bar{F}(\bar{H},\bar{h}_p,\bar{z}) \propto 16^{\Delta_p}.
\label{simplifiedblock}
\end{equation}
An expression for the averaged OPE coefficients squared is known \cite{Das:2017cnv} (see also \cite{Cardy:2017qhl, Collier:2019weq}). When the exchange primary is heavy it reduces to
\begin{equation}
    \overline{|f_{O_HO_HO_{h_p}}|^2}  \approx \frac{16^{-\Delta_p}}{\sqrt{2}} \left(\frac{12\Delta_p}{c-1}-1\right)^{\nu}e^{-2\pi\sqrt{\frac{c-1}{12}\left(\Delta_p-\frac{c-1}{12}\right)}}.
\label{averagedOPEcoefficients}
\end{equation}
where $\nu$ is given by
\begin{equation}
    \nu = 4\Delta -\frac{c+1}{4}
\end{equation}
Combining \eqref{averagedOPEcoefficients} with \eqref{simplifiedblock} and inserting this combination in the integral \eqref{crosschannelintegral} results in
\begin{equation}
    Q(z_i,h,0)F(\sigma^2)\bar{F}(\sigma^2) = \int_0^{\infty} \frac{d\Delta_p}{\sqrt{2}}\; Q(z_i,h,\Delta_p) \left(\frac{12\Delta_p}{c}-1 \right)^{8H -c/4} e^{-2\pi\sqrt{\frac{c}{12}(\Delta_p-\frac{c}{12})}}.
\label{crossedchannelintegral2}
\end{equation}
Setting the $\mathcal{O}(c^0)$ ratios
\begin{equation}
    \delta \equiv \frac{6}{c}H, \;\;\;\; \delta_p = \frac{3}{c}\Delta_p,
\end{equation}
and inserting these into \eqref{crossedchannelintegral2} results in
\begin{equation}
    Q(z_i,h,0)F(H,0;\sigma^2) \propto \int_0^{\infty} d\delta_p \; Q(z_i,h,\delta_p) e^{c \gamma (\delta_p, \delta)},
\label{crossedchannelintegral3}
\end{equation}
where $\gamma(\delta_p,\delta)$ is an $\mathcal{O}(c^0)$ function given by
\begin{equation}
    \gamma(\delta_p,\delta) =  \left(\frac{4}{3}\delta -\frac{1}{4}\right)\log(4\delta_p-1) - \frac{\pi}{6}\sqrt{4\delta_p-1}.
\end{equation}
Since $Q(z_i,h,\delta_p)$ is an $\mathcal{O}(c^0)$ function at large $c$ the integral on the right-hand side of \eqref{crossedchannelintegral3} is sharply-peaked and dominated by the saddle-point of the function $\gamma(\delta,\delta_p)$. The value of $\delta_p^*$ that maximizes $\gamma(\delta_p,\delta)$ is easily solved and given by
\begin{equation}
    \delta_p^{*} = \frac{1}{4\pi^2}\left((3-16\delta)^2+\pi^2\right).
\label{criticaldelta}
\end{equation}
The conclusion is that in the cross-channel something similar happens as was observed in the computation for the S-channel identity block, the data of the initial state gets projected down to a single dominant conformal family. In this case the conformal family that dominates the integral over all exchange operators.

It would seem natural to conjecture that in the case of the state $|V_2\rangle$, constructed out of two heavy operators, that there is a relationship between the Floquet exponent experienced by the light operators and the external scaling dimension of the heavy operators $H$ (or equivalently $\delta$). In section \ref{thermodynamicsandorbits} it was noted in equation \eqref{effectivestresstensor} that a state with Floquet exponent $|\alpha|$ effectively corresponds to the stress-tensor expectation value of a primary state with holomorphic scaling dimension $H_{\text{eff.}}=c(1+|\alpha|^2)/24$. Assuming for simplicity that the dominant T-channel block is scalar and identifying $\delta_p^* = 6H_{\text{eff.}}/c$ leads to the following conjectured relationship between the Floquet exponent and the scaling dimension of the external heavy operators
\begin{equation}
    |\alpha|^2 = \frac{1}{\pi^2}\left(3-\frac{96}{c}H\right)^2, \;\;\;\; H>\frac{c}{32},
\label{monodromyconjecture}
\end{equation}
where it is expected that the regime of validity for $H$ is lower bounded by the root of the above expression. Alternatively, by applying equation \eqref{temperature} a conjecture for the coarse-grained temperature of the state $|V_2\rangle$ can be obtained
\begin{equation}
    T_{|V_2\rangle} = \frac{1}{\pi^2}\left(3-\frac{96}{c}H\right), \;\;\;\; H>\frac{c}{32}.
\label{effectivetemperature2}
\end{equation}
There exists a simple consistency check for the relationship between the S-channel identity block and the dominant saddle \eqref{criticaldelta}. It was noted that $H_{\text{eff}}$ in \eqref{effectivestresstensor} was bounded from below by $H_{\text{eff}}\geq c/24$, it was commented that the value $H_{\text{eff}}$ that saturates this bound corresponds to a minimal mass BTZ black hole in an AdS$_3$ holographic dual \cite{Fitzpatrick:2015zha, fitzpatrick2016conformal}. The value $H_{\text{eff.}}=c/24$ corresponds to $\delta_p^* =1/4$ in \eqref{criticaldelta}. This value is the minimal value for $\delta_p^*$ and is attained when $\delta =3/16$, i.e. $H=c/32$. It was found numerically in \cite{Banerjee:2024qgg} that $\delta=3/16$ is the exact value where the four identical heavy operator S-channel identity block undergoes a transition mimicking a bulk black hole phase transition.

\section{Summary}
The subject of this paper was the normalized expectation value of two light operators on a heavy state $|V\rangle$,
\begin{equation}
    \frac{\langle V|O_L(z_1)O_L(z)|V\rangle}{\langle V|V\rangle}.
\end{equation}
It was assumed that there exists an S-channel OPE-expansion where the expectation value is dominated by the identity block exchange. Using the monodromy method the factor that depends on $z_i$ was given by \eqref{lightsector}
\begin{equation}
    Q(z_1,z_2)  = \frac{1}{\left(\rho_+(z_1)\rho_-(z_2) - \rho_+(z_2)\rho_-(z_1)\right)^{2h}}.
\label{lightsector6}
\end{equation}
A remarkably compact expression, despite that it depends on a set of function $\rho_{\pm}(z)$ which are very difficult to determine in practice. The functions $\rho_{\pm}(z)$ are the linearly independent solutions of \eqref{leadingFuchsian}, i.e. the leading part of the null-vector decoupling equation 
\begin{equation}
    \rho_{\pm}'' + \frac{6}{c}T_H(z)\rho_{\pm}=0.
\label{leadingFuchsian2}
\end{equation}
The expression \eqref{lightsector6} has a couple of nice properties, for one it is by construction completely independent of choice of basis for the solutions of \eqref{leadingFuchsian2}. Secondly, in the case of a primary state, when the solutions are known \eqref{lightsector6} reduces to known expressions \cite{Fitzpatrick:2015zha, Banerjee:2018tut}, thirdly in the limit where $z_2\rightarrow z_1$ it reduces to the conformal two-point function.

While the solutions of the ODE \eqref{leadingFuchsian2} are generally unknown it was shown in section \ref{Lorentziantime} that their branch structure has a familiar physical characteristic. If the locations of the light operators are analytically continued to Lorentzian time than it was found that after each ``light-crossing time" on the compact spatial dimension a number of branch discontinuities is picked up that exactly corresponds to the crossings of the light operators of the lightcones of the heavy operators in the initial state. This is in line with the expectations of \cite{Kundu:2025jsm}. 

Finally, a small puzzle was presented. It was found that the only information of the state $|V\rangle$ that entered in the time-evolution of the light operators was the derived function \eqref{Hillpotential}
\begin{equation}
\mathcal{V}(t) = 1-\frac{24}{c}e^{4it}T(t).
\end{equation}
It was shown that this function has the required properties to be considered part of the the Virasoro coadjoint vector
\begin{equation}
    (\mathcal{V}(t)dt^2, 12\pi) \in \overline{\text{Diff}}^*(S^1)/\mathcal{S}^1.
\end{equation}
The puzzle is that a priori $|V\rangle$ is a linear combination of many states of many different conformal families, whereas $(\mathcal{V}(t)dt^2, 12\pi)$ is an element of a single Virasoro coadjoint orbit, i.e. the classical limit of a single Virasoro representation. 

In section \ref{conformalbootstrap} the heavy state $|V_2\rangle$ was considered. This is a special case where the heavy state is built out of two heavy operators. This results in a CFT problem that can be schematically indicated as the HHHHLL problem. In this case it was found that the above-mentioned projection to a single conformal representation has a direct manifestation in the conformal cross-channel of the heavy sector. From known expressions form the conformal block \cite{Zamolodchikov:1987avt} and known asymptotic expression for the averaged OPE coefficients \cite{Das:2017cnv,Cardy:2017qhl,Collier:2019weq} it was found that the sum over cross-channel heavy blocks is sharply-peaked at a dominant term in the sum. By equating the location of the saddle-point with the data of the orbit a conjecture was presented in  equation \eqref{effectivetemperature2} for a relationship between the external scaling dimension of the heavy operators in $|V_2\rangle$ and the effective temperature experienced by light operators on this state $\langle V_2| O_LO_L|V_2\rangle$.

\section{Discussion and outlook}
A generic expression was constructed for the expectation value of two light operators on a heavy state. The heavy state was constructed in a way that made it manifestly time-dependent. It was found that Floquet's theorem allows for a class of states that are controlled by exponential behavior. This suggests a natural coarse-graining scale. If the time separation of the two light operators is significantly larger than the length scale of the compact spatial dimension than effectively the periodic details of the expectation value get washed out with respect to the overall exponential behavior. This Floquet exponent is a representation theoretic object that sets an effective temperature. The reduction of all the information of the initial state down to just one number is suggestive of a kind of 2d CFT no-hair theorem.

The power of large central charge is that it reduces the, in principle, extremely complicated strongly-coupled problem of computing the expectation value to an ODE problem with what looks like a single degree of freedom. This is still a problem with two layers of complexity, first solving the ODE for four heavy operators or more is very challenging outside of a few sporadic cases, but secondly the information of the CFT state is not contained in the solution of the ODE but in the potential of the ODE, which is itself extremely complicated to obtain\footnote{As was seen in section \ref{thermodynamicsandorbits} though, if given a basis of solutions one could reconstruct the potential from \eqref{reconstructedpotential}}. In fact, as was seen in the discussion of the monodromy method, knowing the residues of the simple poles of the stress-tensor expectation value is equivalent to knowing the semi-classical conformal block. It is the overall structure of the stress tensor expectation value (fixed by the Virasoro Ward identity) along with the compactness of the spatial dimension that allows for enough structure to make meaningful physical statements using Floquet's theorem.

From the perspective of hyperbolic geometry, the mapping \eqref{quasisymmetricmap} plays the role of a boundary value of a quasiconformal map (i.e. a quasisymmetric map). These maps can be considered as elements of Teichm\"uller space \cite{Ahlfors1966,Scarinci:2011np}. Since $\alpha$ is non-zero the Schwarzian derivative of \eqref{quasisymmetricmap} has an additional term proportional to $|\alpha|^2$. Naively this looks like the four-punctured sphere is universally covered by a disk with a hole cut out. It would be interesting to see if the conjectured relationship between the data of the monodromy around single singular points and the monodromy around a loop surounding two singular points \eqref{monodromyconjecture} has a manifestation in the Riemann surface literature.


It is interesting to compare the situation presented here with Floquet CFT \cite{Wen:2018agb}. The difference in this case is that in Floquet CFT the Hamiltonian is time-dependent, i.e. the model itself is periodically driven. Contrast this with the situation presented in this article where it is the state that is time-dependent and drives the expectation values of light local observables. Despite this difference a similar mechanism seems to drive thermalization in both cases. 

Finally, the analysis above focused exclusively on adding only two light operators. The derivation in section \ref{monodromymethod} extends equally well to any number of even light operators as long as one considers an OPE channel where the pairs are contracted to the identity operator. In this case the analytic continuation to Lorentzian would be richer in its caual structure. It would be interesting to extend the analysis to four light-operators in order to consider the out-of-time-ordered correlator of light operators. This would open the door for computing diagnostics of quantum chaos.

\section*{Acknowledgements}
This work was in large part inspired by an earlier publication, the author is therefore especially grateful for the work of Souvik Banerjee, Jan-Willem Brijan. In addition this work greatly benefited from various discussions with various people, in particular, Sungjay Lee, Kyriakos Papadodimas, Tom\'as Proch\'azka, Shiraz Minwalla and Joris Raeymaekers. This work was supported by a KIAS Individual Grant (PG100401) at Korea Institute for Advanced Study.

\appendix

\section{The conformal invariance of the Fuchsian equation}
\label{ODEtransformation}
We will check that the general form of the Fuchsian equation is preserved under analytic maps of the Riemann sphere. In the process we can derive a transformation rule for the field $\psi(z)$ and the stress-tensor $T(z)$. We will start by assuming the Fuchsian equation
\begin{equation}
    \rho''(z) + \frac{6}{c}T(z)\rho(z) =0. 
\end{equation}
Perform the following transformation
\begin{equation}
    z \rightarrow w(z),
\end{equation}
then the differential equation takes the form
\begin{equation}
    \frac{d^2}{dz^2}\rho(w(z)) + \frac{6}{c}T(z(w)) \rho(w(z)) = 0. 
\end{equation}
By defining $d\rho(w)/dw = \rho'(w)$ and $dw/dz = w'$ we find from the chain rule that
\begin{equation}
  \rho''(w)w'^2 + \rho'w'' + \frac{6}{c}T(z) \rho(w) = 0.
\end{equation}
Dividing by $w'^2$ gives
\begin{equation}
    \rho''(w) + \frac{w''}{w'^2}\rho(w) + z'(w)^2 \frac{6}{c}T(z)\rho(w)=0
\label{transformedFuchsian}
\end{equation}
We can use the ODE transformation rule provided in \cite{MagnusWinkler} which states that if an ODE is of the form
\begin{equation}
    \frac{d^2z}{dx^2} + a(x) \frac{dz}{dx} + b(x) z(x) = 0,
\end{equation}
it can be transformed to the form
\begin{equation}
    \frac{d^2y}{dx^2} + Q(x)y(x) =0,
\end{equation}
by the identifications
\begin{equation}
    y(x) = e^{\frac{1}{2}A(x)}z(x), \;\;\;\;\; \frac{dA}{dx} = a(x),
\end{equation}
and
\begin{equation}
    Q(x) = -\frac{1}{2}\frac{da}{dx} - \frac{1}{4}a(x)^2 + b(x).
\end{equation}
Going back to \eqref{transformedFuchsian} we see that we first have to invert the coefficient in front of the first order derivative
\begin{equation}
    \frac{w''}{w'^2} = z'^2 \frac{d}{dz} \frac{1}{z'} = z'^2 \frac{1}{z'} \frac{d}{dw} \frac{1}{z'} = -\frac{z''}{z'}.
\end{equation}
From this we see that our analogue of $a(w)$ in the transformation rule of \cite{MagnusWinkler} is given by
\begin{equation}
    a(w) = - \frac{z''}{z'} \;\;\; \implies \;\;\; A(w) = -\log(z').
\end{equation}
From this we read off that equation \eqref{transformedFuchsian} can be recast into the form
\begin{equation}
    \tilde{\rho}''(w) + \frac{6}{c} \tilde{T}(w) \tilde{\rho}(w) = 0,
\end{equation}
where 
\begin{equation}
    \tilde{\rho}(w) \equiv e^{-\frac{1}{2}\log(z')}\rho(w) = \frac{1}{\sqrt{z'}}\rho(w),
\end{equation}
and
\begin{align}
    & \frac{6}{c}\tilde{T}(w) \equiv -\frac{1}{2} \frac{d}{dw} \left(-\frac{z''}{z'}\right) -\frac{1}{4}\left(\frac{z''}{z'}\right)^2 + \frac{6}{c}T(z(w))z'^2 \nonumber\\ 
    & = \frac{6}{c}\left(z'^2 T(z(w)) + \frac{c}{12}\{z,w\}\right). 
\end{align}
From this we can see the following. The potential $T(z)$ transforms as a quadratic differential, i.e. as a 2d CFT stress tensor. Secondly the solution $\psi(z)$ transforms as a conformal primary with scaling dimension -1/2. Note that this is consistent with the monodromy method, where the function $\psi(z)$ has a physical interpretation as the expectation value of a degenerate ghost field with scaling dimension 
\begin{equation}
    h = \frac{1}{16}\left(5-c+\sqrt{25-26c+c^2}\right).
\end{equation}
When $c\gg1$ this asymptotes to
\begin{equation}
    h\overset{c\rightarrow\infty}{\rightarrow} -\frac{1}{2} + \mathcal{O}(1/c).
\end{equation}
So in the regime where the monodromy method is valid $\psi(z)$ does indeed transform as a primary field with scaling dimension -1/2.

\section{Virasoro coadjoint orbits}
\label{vcoadorbits}
This appendix will provide a very superficial introduction to Virasoro coadjoint orbits. The goal is just to provide enough necessary terminology and intuition to follow section \ref{relationshiptoorbits}. For classic works the reader is referred to \cite{Kirillov1976,WittenCoadjointOrbits, Segal1981, LazPan75} and for contemporary review \cite{Balog:1997zz,Cotler:2018zff}.  

The upcoming discussion can be summarized as follows, the Virasoro group will be considered as an extended version of the group of orientation-preserving diffeomorphisms of the circle. By taking inspiration from the Virasoro charge algebra a bracket, a dual space and a bilinear form will be constructed on the extended algebra. The coadjoint action will be constructed by computing the equivalent of the bracket on the dual space. The coadjoint action gives the tangent space of a coadjoint action up to a number of stable directions. In the process of constructing these stable directions Hill's equation makes a natural appearance.

As a warm-up, the group under consideration is the group of all orientation-preserving difeomorphisms of the circle Diff$(S^1)$. The elements $s(t)$ of Diff$(S^1)$ satisfy:
\begin{equation}
    s(t+\pi) = s(t) +\pi, \;\;\;\; s'(t)>0.
\end{equation}
The Lie algebra of this group is the set of all infinitessimal deformations of these maps, i.e. the vector fields on the circle $f(t)\frac{d}{dt}$. These vector fields possess a basis decomposition in terms of Fourier modes
\begin{equation}
    l_n = \frac{i}{2} e^{2int}\frac{d}{dt},
\end{equation}
which results in the Lie bracket algebra
\begin{equation}
    [l_n,l_m] = (n-m)l_{n+m},
\end{equation}
which can be recognized as the Witt algebra.

The Virasoro algebra is the central extension of the Lie bracket algebra of vector fields on $f(t)\frac{d}{d t}$ on $S^1$. This central extension will be denoted by $\overline{\text{diff}}(S^1)$. The central extension can be constructed by promoting the vector fields to doublets by appending an additional number 
\begin{equation}
    (f(t)\frac{d}{dt}, a) \in \overline{\text{diff}}(S^1), \;\;\;\; a \in\mathbb{R}.
\end{equation}
The appropriate algebra that acts on this central extension can be constructed from the Virasoro charge algebra. Consider the charges associated with infinitesimal complex diffeomorphisms $\epsilon(z)\frac{d}{dz}$ of the complex plane
\begin{equation}
    Q_{\epsilon}[T] = \frac{1}{2\pi i} \oint dz \; \epsilon(z)T(z),
\end{equation}
the associated charge algebra is
\begin{equation}
    [Q_{\epsilon_1}[T],Q_{\epsilon_2}[T]] = -Q_{\epsilon_2}[\delta_{\epsilon_1}T].
\label{chargebracket}
\end{equation}
The effect of an infinitessimal shift in the complex coordinate $z$ has an effect on the stress tensor given by
\begin{equation}
    \delta_{\epsilon}T(z) = -2T(z)\partial \epsilon(z) - \epsilon(z) \partial T(z) - \frac{c}{12}\partial^3 \epsilon(z).
\end{equation}
Inserting this expression in \eqref{chargebracket} and performing a number of integrations by parts leads to the bracket
\begin{equation}
    [Q_{\epsilon_1}[T],Q_{\epsilon_2}[T]] = Q_{[\epsilon_1,\epsilon_2]}[T] - \frac{c}{24\pi i }\oint dz\; \frac{1}{2}(\partial \epsilon_2\partial^2 \epsilon_1^2 - \partial \epsilon_1\partial^2 \epsilon_2^2),
\end{equation}
where a number of integrations by parts have been performed on the inhomogeneous term in order to make it manifestly anti-symmetric in $\epsilon_i(z)$ and where $[\epsilon_1,\epsilon_2]$ is the Lie bracket on $\epsilon_i\frac{d}{dz}$. This provides inspiration for defining the following bracket on the doublet
\begin{equation}
    [(f(t)\frac{d}{dt},a),(g(t)\frac{d}{dt},b)] = \left((fg'-g'f)\frac{d}{dt},\, -\frac{1}{48\pi}\int_0^{\pi} g'f''-f'g''dt\right).
\label{VirasoroBracket}
\end{equation}
Setting the following set of  basis generators for the algebra
\begin{equation}
    L_n = (l_n, 0), \;\;Z= (0,-i)
\end{equation}
We find that the extended Lie algebra is given by
\begin{align}
    & [L_n,L_m] = (n-m)L_{n+m} + \frac{Z}{12} n^3\delta_{n+m,0},\\
    & [L_n,Z] = 0.
\end{align}
So we confirm that this is a central extension of the Witt algebra with a central element $Z$ and associated central charge. The Virasoro algebra here as the slightly unconventional form of \cite{WittenCoadjointOrbits}, as stated in there, the conventional Virasoro algebra can be obtained by shifting the $L_0$ generator by a constant.

\subsection{Virasoro coadjoint action}
Since the components of the Lie algebra consist of a vector field and a number, i.e. $(f(t)\frac{d}{dt}, a)$ there exists a natural dual space with associated inner product. Consider the space of doublets consisting of a quadratic differential and a number
\begin{equation}
    (G(t)dt^2, a) \in \overline{\text{diff}}(S^1)^{*},
\end{equation}
there exists a natural inner product between elements of $\overline{\text{diff}}(S^1)^{*}$ and $\overline{\text{diff}}(S^1)$ given by
\begin{equation}
    \langle (G(t)dt^2,a), \; (f(t)\frac{d}{dt},b)\rangle \equiv \int_0^\pi fG \; dt  \; +ab.
\end{equation}
Define the adjoint action of the algebra on itself through means of the commutator 
\begin{equation}
    \text{ad}_u(v)=[u,v],
\end{equation}
explicitly, this takes the form of equation \eqref{VirasoroBracket}
\begin{equation}
    \text{ad}_{(f(t)\frac{d}{dt},a)}((g(t)\frac{d}{dt},b)) = \left((fg'-g'f)\frac{d}{dt},\, -\frac{1}{48\pi}\int_0^{\pi} g'f''-f'g''dt\right).
\end{equation}
The coadjoint action ad$^{*}$ is the action on the dual space such that it reproduces\footnote{up to a sign, which is conventional to ensure the pairing $\langle G,f\rangle$ is invariant under a simultaneous (co)adjoint transformation of the left- and right-entry \cite{WittenCoadjointOrbits}} the action of the adjoint action
\begin{equation}
    \langle \text{ad}^{*}_{(g(t)\frac{d}{dt},a)}(G(t)dt^2,c),\; (f(t)\frac{d}{dt},b)\rangle = -\langle (G(t)dt^2,a),\; \text{ad}_{(g(t)\frac{d}{dt},c)} (f(t)\frac{d}{dt},b)\rangle
\label{adjointocoadjoint}
\end{equation}
By performing a number of integrations the right hand-side of \eqref{adjointocoadjoint} can be reorganized into
\begin{equation}
      \langle \text{ad}^{*}_{(g(t)\frac{d}{dt},a)}(G(t)dt^2,c),\; (f(t)\frac{d}{dt},b)\rangle = \int_0^{\pi} \left(2f'G+G'f +\frac{c}{24\pi}f'''\right)g(t) \;dt.  
\end{equation}
From which the coadjoint action can be read off
\begin{equation}
   \boxed{\text{ad}^{*}_{(f(t)\frac{d}{dt},a)}(G(t)dt^2,c) = \left(2f'G+G'f +\frac{c}{24\pi}f''', 0\right).} 
\label{coadjointactionappendix}
\end{equation}
Which, up to a sign, can be recognized as the transformation rule of the 2d CFT stress tensor under infinitessimal conformal transformations. Conversely, we find that coadjoint vectors transform as CFT stress tensor under the action of the Virasoro algebra.

\subsection{Virasoro coadjont orbits and Hill's equation}
As was shown in the previous section, the set of pairings consisting of a quadratic differential and a number $G=(G(t)dt^2, c_0)$ form a coadjoint vector of the Virasoro algebra. The coadjoint action that acts on these coadjoint vectors is given in \eqref{coadjointactionappendix}. One can consider a Virasoro coadjoint orbit
\begin{equation}
    \mathcal{O}_{G_0} = \{\text{ad}^*_g \,G_0\; |\, \forall g \in \overline{\text{Diff}}^*(S^1)\},
\end{equation}
a manifold that consists of a continuous set of all Virasoro coadjoint vectors that are related to a given reference coadjont vector $G_0 = (G_0(t)dt^2, a_0)$ by exponentiating the coadjoint action to a group action.

Before considering the manifold structure of the orbit consider first the local tangent space of the orbit at the reference vector $G_0$.
Given a fixed reference coadjoint vector $G_0$ there will be in principle be a set $h_0$ of stabilizer Lie algebra elements that annihilate the reference point, i.e.
\begin{equation}
    \text{if} \;\;\;\; (f(t)\frac{d}{dt},b) \in h_0 \;\;\; \text{then} \;\;\; \text{ad}^*_{(f(t)\frac{d}{dt},b)} G_0 = 0.
\end{equation}
If we mod out the stabilizer algebra then we find that $\overline{\text{diff}}(S^1)/h_0$ is the tangent space of such an orbit at the point $G_0$. All coadjoint orbits are homogeneous\footnote{In non-technical terms: any coadjoint vector can be reached from any other coadjoint vector by an element of the Virasoro group. In technical terms: the Virasoro group acts transitively on the quotient space $\overline{\text{Diff}}(S^1)/H$.} spaces \cite{Kirillov1976}, so the subscripts can be dropped, i.e. $(G(t)_0dt^2, c_0)\rightarrow (G(t)dt^2, c)$ and $h_0\rightarrow h$. In principle the full orbits can be constructed by exponentiating the tangent vector fields resulting in the homogeneous space $\overline{\text{Diff}}(S^1)/H$, where $H$ is the Lie group obtained by exponentiating the algebra $h$ \cite{Kirillov1976}. 

The possible stabilizer algebras have been classified. Take the reference point $G$. The condition that $f_0 = (f_0(t)d/dt, a)$ is an element of the stabilizer algebra is that
\begin{equation}
    2f_0'G+G'f_0 +\frac{c}{24\pi}f_0''' = 0.
\end{equation}
Locally, the solutions to the stabilizer can be expressed in terms of the solutions of a 2nd order linear ODE\footnote{Specifically an equation of Hill type \cite{Balog:1997zz,MagnusWinkler}}. If the two linearly independent functions $\psi_{\pm}$ solve the equation
\begin{equation}
    \psi_{\pm}''(t) + \frac{12\pi}{c}G(t)\psi_{\pm}(t) = 0,
\label{Hillseqstabilizer}
\end{equation}
then the three functions
\begin{equation}
    f_1(t) = \psi_+^2, \;\;\; f_2(t)=\psi_-\psi_+, \;\;\; f_3(t)=\psi_-^2
\end{equation}
are the three independent solutions of the stabilizer equation for $(G(t)dt^2, c)$, i.e.
\begin{equation}
    2f_i'G+G'f_i +\frac{c}{24\pi}f_i''' = 0.
\end{equation}
These solutions are only local though, they do not necessary satisfy the periodicity condition required to be considered vector fields on $S^1$. A proof in \cite{WittenCoadjointOrbits} shows that for any given $G(t)$ there can only be either 1 or 3 periodic solutions to the stabilizer equation. This discussion also demonstrates that there is a direct link between Hill's equation and the classification of Virasoro coadjoint orbits.

\bibliography{references}
\bibliographystyle{JHEP}

\end{document}